\documentclass[preprintnumbers,11pt,a4paper,oneside]{article}

\usepackage{contour}
\usepackage{xcolor}
\usepackage{float}
\usepackage{mathrsfs}
\contourlength{0.05em}

\usepackage{jheppub}
\usepackage{color}
\usepackage{graphicx}
\usepackage{wrapfig,enumerate,slashed}

\usepackage{footmisc}
\usepackage{amsmath}
\usepackage{wasysym} 
\usepackage{graphicx}
\usepackage{booktabs}
\usepackage{subcaption}
\usepackage{color}
\usepackage{comment}
\usepackage{appendix}
\usepackage{slashed}
\usepackage{natbib}

\newcommand{\be}{\begin{equation}}
\newcommand{\ee}{\end{equation}}
\newcommand{\beq}{\begin{equation}}
\newcommand{\eeq}{\end{equation}}

\newcommand{\bee}{\begin{eqnarray}}
\newcommand{\eee}{\end{eqnarray}}
\newcommand{\beeq}{\begin{equation}}
\newcommand{\eeeq}{\end{equation}}

\DeclareRobustCommand{\Sec}[1]{Sec.~\ref{#1}}

\DeclareRobustCommand{\Fig}[1]{Figure~\ref{#1}}

\DeclareRobustCommand{\Eq}[1]{Eq.~(\ref{#1})}

\usepackage{tikz}
\usepackage{orcidlink}

\makeatletter
\gdef\@fpheader{}
\makeatother
\begin{document}

\title{Quantum-Inspired Tensor Network Autoencoders for Anomaly Detection: A MERA-Based Approach}

\author[a]{Emre Gurkanli$^*$\,\orcidlink{0000-0002-9543-1086}\note{Corresponding author.}}
\author[b,c]{and Michael Spannowsky$^*$\,\orcidlink{0000-0002-8362-0576}}

\emailAdd{egurkanli@sinop.edu.tr}
\emailAdd{michael.spannowsky@kit.edu}

\affiliation[a]{\vspace{0.1cm} Department of Physics, Sinop University, Turkiye}
\affiliation[b]{\vspace{0.1cm} Institute for Theoretical Physics, Campus S\"ud, Karlsruhe Institute of Technology (KIT), D-76128 Karlsruhe, Germany}
\affiliation[c]{\vspace{0.1cm} Institute for Quantum Materials and Technologies, Karlsruhe Institute of Technology, Karlsruhe 76131, Germany}

\abstract{
We investigate whether a multiscale tensor-network architecture provides a useful inductive bias for reconstruction-based anomaly detection in collider jets. Jets are produced by a branching cascade, so their internal structure is naturally organised across angular and momentum scales. This motivates an autoencoder that compresses information hierarchically and can reorganise short-range correlations before coarse-graining. Guided by this picture, we formulate a Multiscale Entanglement Renormalization Ansatz (MERA)-inspired autoencoder acting directly on ordered jet constituents. To our knowledge, a MERA-inspired autoencoder has not previously been proposed, and this architecture has not been explored in collider anomaly detection.

We compare this architecture to a dense autoencoder, the corresponding tree-tensor-network limit, and standard classical baselines within a background-only reconstruction framework. The paper is organised around two main questions: whether locality-aware hierarchical compression is genuinely supported by the data, and whether the disentangling layers of MERA contribute beyond a simpler tree hierarchy. To address these questions, we combine benchmark comparisons with a training-free local-compressibility diagnostic and a direct identity-disentangler ablation. In the nominal benchmark, the MERA-inspired model attains a higher mean AUC than the dense autoencoder at all three latent dimensions while using roughly one-third of its trainable parameters. The locality-preserving multiscale structure is well matched to jet data, and, at the strongest compression bottleneck, the trained disentanglers move substantially away from the identity and are accompanied by a positive mean performance shift. Overall, the study supports locality-aware hierarchical compression as a useful and parameter-efficient inductive bias for reconstruction-based jet anomaly detection in this benchmark setting.
}

\maketitle


\section{Introduction}

\noindent
The Standard Model (SM) gives an excellent description of collider data, but it does not explain several basic observations, including dark matter, neutrino masses, and the baryon asymmetry of the Universe. The search for physics beyond the SM, therefore, remains a central goal of the LHC programme. In practice, however, most searches are optimised for specific benchmark signatures. This is a powerful strategy when the signal is known in advance, but it can reduce sensitivity to unexpected phenomena that do not fit a standard search template. For this reason, there is increasing interest in search strategies that rely less strongly on a detailed signal hypothesis and instead look more broadly for deviations from the dominant SM background~\cite{Belis:2024anomaly,Kasieczka:2021lhc}.

Anomaly detection provides one such framework. The basic idea is to learn what typical background events look like and then assign a large anomaly score to events that are poorly described by that learned background model~\cite{Belis:2024anomaly}. In collider physics, this programme has grown rapidly in recent years, supported by community benchmarks and systematic comparisons that clarified both the promise and the limitations of model-agnostic searches~\cite{Kasieczka:2021lhc,Fraser:2021lxm}. A recurring lesson is that the choice of data representation and the built-in structure of the model are crucial: an anomaly detector should be flexible enough to respond to a wide class of signals, while remaining robust against ordinary fluctuations of the background.

Among the many possible approaches, reconstruction-based methods are especially attractive because they are conceptually simple and easy to compare across architectures. In these methods, a model is trained only on background events to compress and reconstruct its input. The reconstruction error is then used as the anomaly score, under the expectation that atypical events will be reconstructed less accurately. Autoencoders have therefore become a standard reference point in collider anomaly detection, including for jet-based searches, resonant-search studies, and more specialized adversarial, normalized, or graph-based constructions~\cite{Heimel:2019qcd,Farina:2020deepae,Collins:2021weakunsup,Barron:2021suep,Blance:2019adversarialae,Dillon:2023normalizedae,Atkinson:2022ircgraphae,Finke:2021autoencoders}. At the same time, recent studies have emphasized that reconstruction quality alone is not enough: performance can depend strongly on the input representation, the architecture, and the evaluation protocol~\cite{Fraser:2021lxm,Finke:2021autoencoders}. Related work has also shown that non-neural or weakly supervised alternatives, such as tree-based strategies, can be highly competitive in collider anomaly searches~\cite{Finke:2024treebased}. This makes it especially important to ask what kind of architecture is most naturally matched to the physics of jets.

Jets are collimated sprays of hadrons produced by the fragmentation and hadronisation of energetic quarks or gluons in high-energy collisions. In proton-proton collisions, ordinary QCD jets constitute the dominant background, while jets originating from boosted heavy-particle decays, such as hadronic top-quark decays, typically exhibit richer internal substructure and more complex radiation patterns. These differences make jet substructure a natural setting for anomaly detection studies.

This substructure arises from a branching cascade: a high-energy quark or gluon radiates, those daughters radiate again, and the observable jet is built up through a nested sequence of splittings across several angular and momentum scales~\cite{Larkoski:2020jetsub}. Physically relevant information is therefore not distributed uniformly across constituents. Hard, large-angle splittings determine the coarse prong structure of the jet, while softer and more collinear radiation refines that structure at progressively smaller scales. For anomaly detection, this is precisely the type of organisation one would like a model to capture: signal jets may differ from QCD not only through isolated features, but through how correlations persist or reorganise across scales.

This motivates looking beyond dense autoencoders. A fully connected autoencoder is flexible, but it treats the input as a largely unstructured vector and must learn any hierarchical organisation from the data alone. In other words, the network is capable of representing multiscale correlations, but it is not encouraged to do so in any particular way. This makes a concrete hypothesis testable: if the ordered jet representation is locally compressible in a way that follows the shower geometry, then a hierarchical architecture should model the background more economically than a dense map. Whether the additional disentangling structure of Multiscale Entanglement Renormalization Ansatz (MERA) is genuinely useful beyond a simpler tree hierarchy is, however, an empirical question rather than an assumption.

We study a multiscale architecture inspired by the Multiscale Entanglement Renormalisation Ansatz (MERA)~\cite{Vidal:2007er,Vidal:2008mera}. MERA was originally developed in quantum many-body physics as a tensor-network construction that alternates coarse-graining with disentangling transformations, allowing relevant correlations to be retained across length scales. Related tensor-network ideas have also been adapted to machine learning, where they offer structured and often parameter-efficient representations of high-dimensional data~\cite{Stoudenmire:2016tnml}. In collider physics, tensor-network-inspired approaches have already been explored for event reconstruction, comparisons between classical and quantum tensor-network circuits, and generative or anomaly-oriented modelling~\cite{Araz:2021zwu,Araz:2022haf,Araz:2022zxk}. For jet analysis, the appeal of MERA is, that its coarse-graining steps resemble the progressive aggregation of information from nearby constituents into larger-scale features, while the disentangling operations are designed to keep short-range correlations from being lost before compression. In a jet context, this means that local geometric structure can be reorganised before information is passed to the next scale, rather than being mixed globally from the start.

To the best of our knowledge, this is the first study to formulate collider anomaly detection with a MERA-inspired autoencoder. Earlier anomaly-detection applications in high-energy physics have considered quantum autoencoders and, more recently, tensor-network models based on matrix product states acting in an externally learned latent space, but not a MERA-inspired encoder--decoder operating directly on ordered jet constituents~\cite{Ngairangbam:2022qaead,Puljak:2025tnad}.

This is why we consider the MERA-type anomaly detection worth studying. If the dominant background really lies on a multiscale manifold shaped by the QCD shower, then a MERA-inspired autoencoder should be able to model that background economically: it should retain the correlations that matter, discard redundant short-distance structure efficiently, and compress the jet representation with fewer parameters than a dense network of comparable reach. The public Top Quark Tagging Reference Dataset, introduced as part of a broad comparison of modern jet-tagging methods, provides a clean and widely used benchmark for testing this idea in a controlled setting~\cite{Kasieczka:2019landscape}. Thus, we turn that motivation into two concrete tests: whether a locality-preserving constituent chain is measurably more compressible at the local tensor-network level, and whether trainable MERA disentanglers improve on the corresponding identity-disentangler limit. The goal of the present work is therefore to investigate whether a hierarchical MERA-inspired architecture provides a useful inductive bias for jet representations in a controlled anomaly-detection benchmark.

To this end, we perform a comparison between a dense autoencoder baseline, a MERA-inspired neural architecture, the corresponding tree tensor network (TTN) limit, and classical unsupervised methods, including principal component analysis (PCA), Gaussian modelling, and Isolation Forest. All methods are trained on background-only QCD jets and evaluated against hadronically decaying top jets using identical inputs, preprocessing, and performance metrics. This common setup allows observed differences to be attributed directly to the underlying representation rather than to differences in data handling or optimisation.

We study the models across several latent dimensions and assess their behaviour through ROC curves, AUC values, local-compressibility diagnostics, and direct architectural ablations. Particular emphasis is placed on identifying which part of the multiscale construction is actually supported by the data.

The rest of the paper is organised as follows. In \Sec{sec:mera_construction}, we introduce the canonical MERA as the physical and architectural motivation for the proposed MERA-inspired autoencoder and develop the mathematical framework underlying its construction. In \Sec{sec:data}, we describe the jet representation, the locality-preserving ordering, and the anomaly-detection setup. In \Sec{sec:models}, we present the concrete MERA-inspired, TTN, dense-autoencoder, and classical baseline models used in the study. In \Sec{sec:results}, we report the nominal benchmark, the local-compressibility and ordering analyses, and the disentangler ablation. Finally, \Sec{sec:conclusions} summarises the main findings and their implications.

\section{Constructing a MERA-Inspired Anomaly Detector}
\label{sec:mera_construction}

\begin{figure}[H]
\centering
\includegraphics[width=0.70\textwidth]{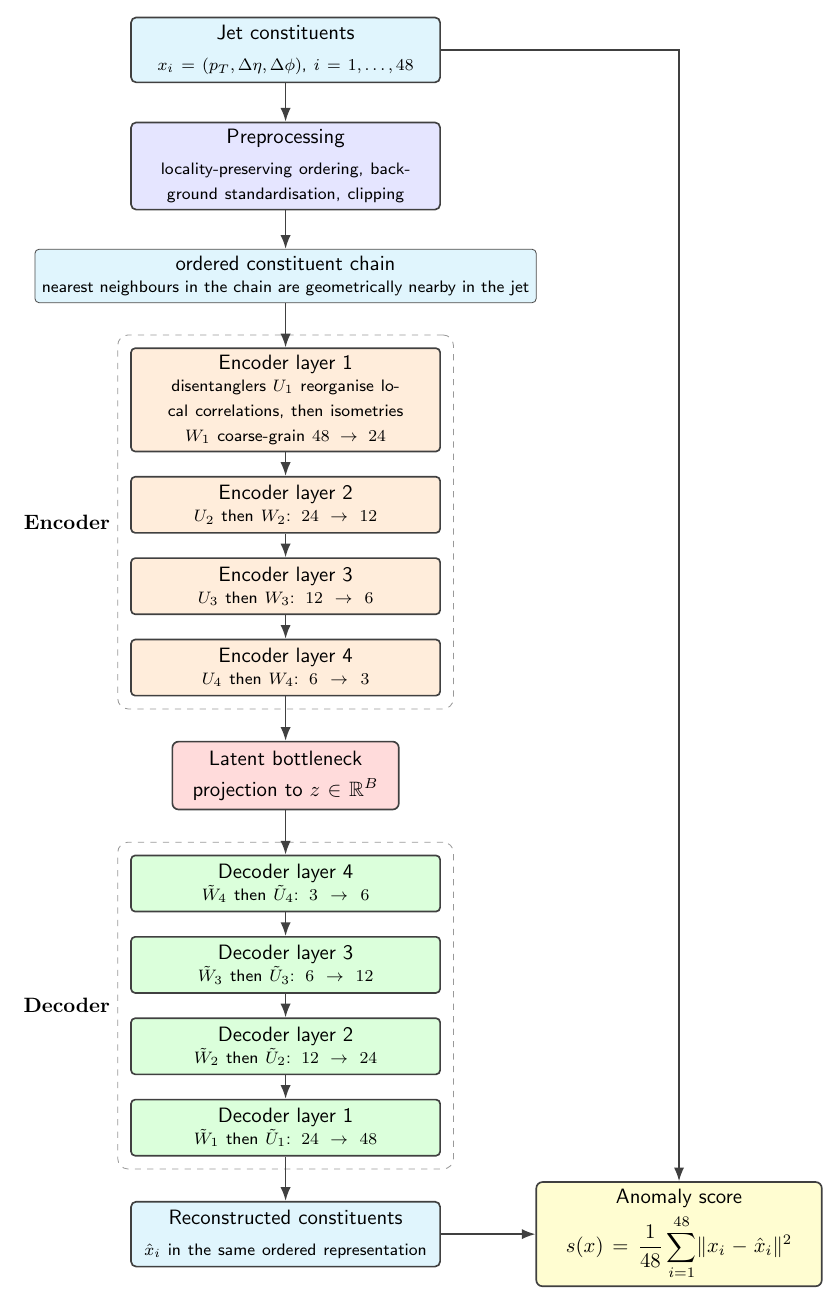}
\caption{Schematic overview of the MERA-inspired neural autoencoder used in this work. After locality-preserving ordering and standardisation, the encoder alternates local mixing maps $U_\ell$ and coarse-graining maps $W_\ell$ recursively across all four layers of the hierarchy $48\rightarrow24\rightarrow12\rightarrow6\rightarrow3$, followed by a projection to a latent vector $z\in\mathbb{R}^B$. In the implemented architecture, $U_\ell$ and $W_\ell$ act successively on the same local pair; this differs from the staggered cross-boundary connectivity of canonical binary MERA. The decoder mirrors the hierarchy with untied tensors $\tilde U_\ell$ and $\tilde W_\ell$ and reconstructs the ordered jet constituents. The anomaly score is computed from the reconstruction error between the original jet and its reconstruction, as indicated by the direct comparison arrow into the score box.}
\label{fig:mera_autoencoder_overview}
\end{figure}

Tensor networks provide structured factorisations of high-order maps into products of low-order tensors. In many-body physics they are used to represent states with constrained correlation structure; in machine learning the same idea has been adapted to supervised classification, multiscale feature extraction, and generative modelling~\cite{Orus:2014intro,Cichocki:2017tnreview,Stoudenmire:2016tnml,Stoudenmire:2018multiscale,Reyes:2021multiscale,Liu:2019unitarytn,Han:2018mpsgen,Cheng:2019ttngen}. Collider applications of tensor-network-inspired methods have also begun to appear, including matrix-product-state event reconstruction and quantum-probabilistic generative or anomaly-detection frameworks~\cite{Araz:2021zwu,Araz:2022haf,Araz:2022zxk}. A reconstruction-based anomaly detector built from MERA combines this tensor-network viewpoint with the standard background-only autoencoder strategy used in collider anomaly detection: one constructs a constrained multiscale encoder--decoder map, fits it only on background jets, and then uses the reconstruction error as the anomaly score. The overall workflow of the proposed MERA-inspired autoencoder is illustrated in Fig.~\ref{fig:mera_autoencoder_overview}.

We emphasise at the outset that the model studied here is a MERA-inspired neural architecture rather than a canonical tensor-network MERA. Canonical binary MERA uses staggered disentanglers acting across the boundaries of the blocks subsequently coarse-grained by isometries. The implemented model instead applies a local mixing map and a coarse-graining map successively to the same constituent pair, introduces nonlinear activations between hierarchical levels, and uses an untied decoder with soft rather than exact structural constraints. We use the term MERA-inspired because the architecture retains the motivating alternation of local correlation reorganisation and scale-by-scale compression. All numerical results below concern this explicitly defined neural architecture and should not be interpreted as results for canonical MERA.

\subsection{From Canonical MERA to a MERA-Inspired Encoder}

In this section, we first describe the canonical MERA construction that motivates the model. The exact finite MERA-inspired neural architecture used in the numerical benchmark, including its modified connectivity and nonlinearities, is specified later in \Sec{sec:models}.

Let a jet be represented by an ordered list of $N$ constituent-level feature vectors,
\begin{equation}
x = (x_1,\dots,x_N),
\qquad
x_i \in \mathbb{R}^{d_0}.
\end{equation}
In the present work, $d_0=3$ and $x_i=(p_T^i,\Delta\eta_i,\Delta\phi_i)$, but the construction is more general. One may either work directly with these site vectors or first apply a local feature map
\begin{equation}
\phi:\mathbb{R}^{d_0}\to\mathbb{R}^{\chi_0},
\qquad
h_i^{(0)}=\phi(x_i),
\end{equation}
where $\chi_0$ is the local site dimension. The full input is then viewed as an element of a tensor-product space,
\begin{equation}
h^{(0)} \in \bigotimes_{i=1}^{N}\mathbb{R}^{\chi_0}.
\end{equation}

\begin{figure}[t]
\centering
\includegraphics[width=0.82\textwidth]{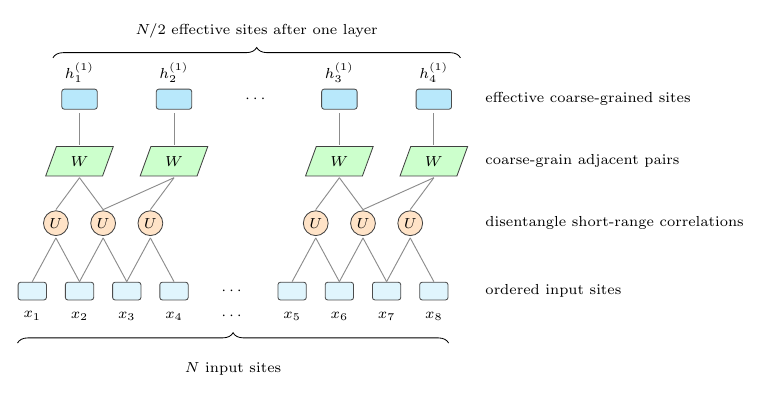}
\caption{Schematic action of one layer in a canonical binary MERA, shown here as the motivation for the proposed architecture. Canonical disentanglers $U$ reorganise short-range correlations across block boundaries before isometries $W$ coarse-grain adjacent pairs into a reduced set of effective sites. The implemented MERA-inspired neural architecture differs in that each $U_j^{(\ell)}$ and $W_j^{(\ell)}$ acts successively on the same local pair, as specified in \Sec{sec:mera_arch}.}
\label{fig:mera_local_motif}
\end{figure}

The essential modelling decision is to interpret the ordered constituents as sites on a one-dimensional hierarchical graph. Once a site ordering has been fixed, a binary MERA encoder acts by alternating local basis changes (the disentanglers) with local projections (the isometries). At layer $\ell$, with $n_\ell$ active sites of dimension $\chi_\ell$, we write schematically
\begin{equation}
h^{(\ell+1)}=
\mathcal{W}^{(\ell)}\,
\mathcal{U}^{(\ell)}\,
h^{(\ell)},
\qquad
\ell=0,\dots,L-1,
\end{equation}
where $\mathcal{U}^{(\ell)}$ is a product of local disentanglers and $\mathcal{W}^{(\ell)}$ is a product of local isometries. In a finite binary MERA-like layout, the local operations may be written as
\begin{equation}
p_j^{(\ell)}
=
h_{a_j}^{(\ell)}\otimes h_{b_j}^{(\ell)}
\in \mathbb{R}^{\chi_\ell^2},
\end{equation}
\begin{equation}
\tilde p_j^{(\ell)} = U_j^{(\ell)}\,p_j^{(\ell)},
\qquad
U_j^{(\ell)} \in \mathbb{R}^{\chi_\ell^2\times \chi_\ell^2},
\end{equation}
\begin{equation}
h_j^{(\ell+1)} = W_j^{(\ell)}\,\tilde p_j^{(\ell)},
\qquad
W_j^{(\ell)} \in \mathbb{R}^{\chi_{\ell+1}\times \chi_\ell^2}.
\end{equation}
Here, the pair labels $(a_j,b_j)$ are determined by the chosen MERA connectivity. In the standard binary construction, the disentanglers act across the boundaries of neighbouring blocks and the isometries then coarse-grain those disentangled blocks into the next layer~\cite{Vidal:2007er,Vidal:2008mera}. The defining structural constraints are
\begin{equation}
(U_j^{(\ell)})^\top U_j^{(\ell)} = I_{\chi_\ell^2},
\qquad
W_j^{(\ell)}(W_j^{(\ell)})^\top = I_{\chi_{\ell+1}},
\end{equation}
so each $U_j^{(\ell)}$ is orthogonal and each $W_j^{(\ell)}$ is an isometry. The action of a single MERA layer, including the disentangling and coarse-graining operations, is illustrated schematically in Fig.~\ref{fig:mera_local_motif}.

After $L$ coarse-graining layers, the remaining top-level tensors are concatenated and mapped to a latent vector
\begin{equation}
z = P\,\mathrm{vec}\!\left(h^{(L)}\right),
\qquad
P \in \mathbb{R}^{B\times (n_L\chi_L)},
\end{equation}
where $B$ denotes the latent dimension of the final representation. In the concrete implementation studied below, this abstract top-level map is specialised to a finite hierarchy whose top tensor has dimension $n_L\chi_L=9$, followed by a final linear projection of width $B$. The encoder is, therefore, a structured map
\begin{equation}
E_\theta:\mathbb{R}^{N d_0}\to\mathbb{R}^{B},
\qquad
z = E_\theta(x),
\end{equation}
with parameter set $\theta=\{U^{(\ell)},W^{(\ell)},P\}$.

\subsection{Decoder and Anomaly Score}

To use MERA for reconstruction-based anomaly detection, the encoder must be paired with a decoder
\begin{equation}
D_\vartheta:\mathbb{R}^{B}\to\mathbb{R}^{N d_0},
\qquad
\hat x = D_\vartheta(z).
\end{equation}
If one insists on an exactly tied construction, the decoder can be built from the transposes of the encoder tensors, since for orthogonal $U$ and isometric $W$ the natural expansion maps are $U^\top$ and $W^\top$. In a more flexible autoencoder realization one may instead introduce an untied decoder,
\begin{equation}
\hat h^{(\ell)} =
\bar{\mathcal U}^{(\ell)}\,
\bar{\mathcal W}^{(\ell)}\,
\hat h^{(\ell+1)},
\end{equation}
with independent parameters $\vartheta=\{\bar U^{(\ell)},\bar W^{(\ell)},\bar P\}$. This generic choice is useful because exact inversion is impossible once information has been discarded by coarse-graining. In \Sec{sec:mera_arch} we adopt this untied option for the concrete benchmark model.

At the conceptual level, the encoder--decoder is trained on background jets and the anomaly score is the reconstruction loss,
\begin{equation}
S(x)=\mathcal{L}_{\mathrm{rec}}\!\left(x,D_{\vartheta^\star}(E_{\theta^\star}(x))\right).
\end{equation}
Large values of $S(x)$ indicate that the jet lies away from the multiscale background manifold learned by the encoder--decoder. This subsection defines only that general reconstruction principle; the exact finite decoder, per-jet loss, soft regularisation terms, and numerical hyperparameters used in the benchmark are specified together in \Sec{sec:mera_arch}.

\subsection{Potential advantages of a MERA approach for anomaly detection and interpretability}

The advantage of MERA over a plain tree tensor network or an unstructured dense map can be made explicit at the level of local compression. Consider a local pair or block vector $p\in\mathbb{R}^{\chi_\ell^2}$ with mean $\mu=\mathbb{E}[p]$ and covariance matrix
\begin{equation}
\Sigma = \mathbb{E}\!\left[(p-\mu)(p-\mu)^\top\right].
\end{equation}
Writing $q=p-\mu$ for the centred local fluctuation, a tree tensor network without disentanglers compresses $q$ by projection onto a low-dimensional subspace,
\begin{equation}
\hat q_{\mathrm{TTN}} = W^\top W\,q.
\end{equation}
By contrast, MERA first rotates the local basis and only then truncates,
\begin{equation}
\hat q_{\mathrm{MERA}} = U^\top W^\top W U\,q.
\end{equation}
The corresponding expected local truncation error is
\begin{equation}
\varepsilon(U,W)
=
\mathbb{E}\!\left[\|q-U^\top W^\top W U q\|_2^2\right]
=
\mathrm{tr}(\Sigma)
-
\mathrm{tr}\!\left(WU\Sigma U^\top W^\top\right).
\label{eq:local_truncation_loss}
\end{equation}
For fixed $U$, the optimal $W$ is obtained by choosing the $\chi_{\ell+1}$ leading eigenvectors of $U\Sigma U^\top$; this is just PCA in the rotated basis. The role of the disentangler is therefore clear: it seeks a basis in which the degrees of freedom that must be retained by the subsequent isometry are as informative as possible. In jet language, this is exactly the mechanism one would like if short-distance angular correlations have to be reorganised before a coarse description is formed. This is the key mathematical reason to expect MERA to outperform a pure tree, and potentially to reach a given reconstruction quality with fewer parameters than a dense autoencoder.

The same argument also explains why the choice of geometry matters. A dense autoencoder compresses the entire vector $x\in\mathbb{R}^{Nd_0}$ using global matrix multiplications. It can represent multiscale structure, but it does not encode where that structure is expected to live. A MERA, by construction, allocates different tensors to different scales and only allows information to propagate upward through a prescribed hierarchy. If that hierarchy is well aligned with the QCD shower, the network spends its capacity on the correlations one expects to be physically relevant, rather than learning them indirectly from scratch. The capacity available at each scale is in turn controlled by the bond dimension $\chi$, which denotes the dimension of the internal tensor indices connecting neighbouring tensors. Physically, larger values of $\chi$ allow more correlations and information to be retained between different parts of the system, at the cost of increased computational complexity.

For moderate bond dimensions, this leads to favourable scaling. If $\chi_\ell\sim \chi$ across layers, then each disentangler carries $\mathcal{O}(\chi^4)$ parameters and each isometry carries $\mathcal{O}(\chi^3)$ parameters, while the total number of local tensors is $\mathcal{O}(N)$ in a binary hierarchy. The resulting parameter count is therefore
\begin{equation}
N_{\mathrm{param}}^{\mathrm{MERA}} = \mathcal{O}(N\chi^4),
\end{equation}
up to the small top tensor. The point is not that this is universally smaller than every dense autoencoder, but that the scaling is linear in the number of sites for fixed bond dimension and that the retained parameters have a clear multiscale interpretation.

\subsection{Mathematical Challenges}

Although the construction above is conceptually natural, designing a MERA for anomaly detection is mathematically non-trivial.

\paragraph{Geometry and site ordering.}
MERA assumes a graph on which locality is defined, whereas a jet is fundamentally an unordered set of constituents. One must therefore specify a map
\begin{equation}
\pi:\{1,\dots,N\}\to\{1,\dots,N\},
\qquad
x \mapsto (x_{\pi(1)},\dots,x_{\pi(N)}),
\end{equation}
that turns the set into an ordered sequence. If this ordering fails to place strongly correlated constituents near each other, then the local disentanglers and isometries will act on the wrong degrees of freedom and the MERA inductive bias is weakened. This is one of the central modelling choices in collider applications, and it is why constituent ordering is not a cosmetic preprocessing step but part of the architecture itself.

\paragraph{Irregular system size.}
The cleanest binary MERA is defined for a dyadic number of sites, $N=2^L$. Jet representations rarely satisfy this exactly, and even a fixed choice such as $N=48$ leads to an irregular top of the hierarchy:
\begin{equation}
48 \rightarrow 24 \rightarrow 12 \rightarrow 6 \rightarrow 3 .
\end{equation}
One must then decide whether to pad to a dyadic size, use mixed-arity tensors near the top, or terminate the hierarchy with an additional dense projection. Each choice changes the balance between exact MERA structure and practical flexibility.

\paragraph{Constrained optimisation.}
The exact parameter spaces are not Euclidean:
\begin{equation}
U_j^{(\ell)} \in \mathrm{O}(\chi_\ell^2),
\qquad
W_j^{(\ell)} \in \mathrm{St}(\chi_{\ell+1},\chi_\ell^2),
\end{equation}
where $\mathrm{O}(m)$ is the orthogonal group and $\mathrm{St}(r,m)$ is the Stiefel manifold\footnote{The Stiefel manifold $\mathrm{St}(r,m)$ is the space of all ordered sets of $r$ mutually orthonormal vectors in $\mathbb{R}^{m}$, or equivalently all $r\times m$ matrices with orthonormal rows. It is the natural parameter space for the MERA isometries, because each $W$ must choose and preserve an orthonormal $r$-dimensional subspace of the local $m$-dimensional block that is retained under coarse-graining.} of $r$ orthonormal rows in $\mathbb{R}^{m}$~\cite{Edelman:1998orthogonality}. Training directly in these constrained spaces is possible, but more involved than ordinary gradient descent. Alternatives include soft penalties of the form used in \Sec{sec:mera_arch}, explicit retractions onto the constraint manifold~\cite{Absil:2008manifolds}, or layerwise constructions based on truncated higher-order singular value decompositions and orthogonal Procrustes updates~\cite{DeLathauwer:2000hosvd,Schonemann:1966procrustes,Batselier:2021meracle}. The practical challenge is that one wants enough freedom to fit the background distribution, while preserving enough orthogonality that the MERA interpretation is not lost.

\paragraph{Non-convex coupling across layers.}
Even the local objective in \Eq{eq:local_truncation_loss} becomes difficult once all layers are trained jointly, because the covariance seen at layer $\ell$ depends on every lower-layer tensor. The optimisation problem is therefore highly non-convex and strongly coupled across scales. Moreover, internal tensor-network gauge freedoms imply that many parameter configurations represent the same global map, producing gauge-induced flat directions in the loss landscape~\cite{Orus:2014intro}. This complicates both optimisation and interpretation of the learned tensors.

\paragraph{No exact inverse after truncation.}
Once $\chi_{\ell+1}<\chi_\ell^2$, the projector $W^\top W$ is not the identity:
\begin{equation}
W^\top W \neq I_{\chi_\ell^2}.
\end{equation}
Therefore, a MERA autoencoder is never an exactly invertible transform after coarse-graining; information has been discarded by construction. For anomaly detection, this is not a bug but the central mechanism, since the model is supposed to keep the dominant background modes and discard the rest. Nevertheless, it means that the decoder must learn a stable approximate inverse of a sequence of projections and rotations, which can be numerically delicate when the retained subspaces vary strongly across layers.

These issues explain why constructing a useful MERA anomaly detector is not equivalent to simply replacing dense layers with tensor contractions. The success of the method depends on simultaneously choosing a meaningful geometry for the data, enforcing a tractable multiscale compression scheme, and training under non-trivial algebraic constraints. The concrete architecture used in our numerical study is one specific realisation of this general construction.

\section{Benchmarking and Data Preprocessing}
\label{sec:data}

\subsection{Benchmark Dataset and Task Definition}

Our study uses the publicly available Top Quark Tagging Reference Dataset, introduced as a common benchmark for comparing jet-substructure and machine-learning methods~\cite{Kasieczka:2019landscape,Kasieczka:2019topdataset}. This benchmark is well-suited to the present analysis for two reasons. First, it is already widely used, which makes the results easy to compare with the existing literature. Second, it provides an environment in which differences between methods can be traced primarily to model design rather than to variations in event generation, detector simulation, or train--test splitting.

The events correspond to proton--proton collisions at a centre-of-mass energy of $\sqrt{s}=14~\text{TeV}$. They are generated with \textsc{Pythia8}~\cite{Sjostrand:2015pythia} and passed through a fast detector simulation based on \textsc{Delphes}~\cite{deFavereau:2014delphes}. Jets are reconstructed with the anti-$k_T$ clustering algorithm~\cite{Cacciari:2008antikt} using radius parameter $R=0.8$. Throughout the analysis, we restrict attention to jets with transverse momentum
\[
550~\text{GeV} \le p_T^{\text{jet}} \le 650~\text{GeV}
\]
and pseudorapidity
\[
|\eta^{\text{jet}}|<2.
\]

A jet is a collimated spray of hadrons produced when an energetic quark or gluon fragments in the detector. In this benchmark, the background sample consists of ordinary QCD jets from dijet production, while the signal sample consists of jets initiated by hadronically decaying top quarks. The top jets are therefore not used here as a literal model of new physics. Rather, they serve as a well-understood example of jets with richer internal structure than the dominant QCD background, making them a useful proxy anomaly when testing reconstruction-based methods. A broader study involving multiple signal classes would be valuable for
assessing anomaly-detection generalisation, but lies beyond the scope of the present work.

The dataset is provided with fixed splits of 1.2M training jets, 400k validation jets, and 400k test jets. We keep these predefined splits unchanged for every method considered in this paper. In the anomaly-detection setup, all models are trained or fitted only on background jets. Signal jets enter only at validation and test time. This is conceptually important: the task is not supervised classification of top versus QCD jets, but the identification of jets that are poorly described by a model of typical background structure.

\subsection{Jet Representation}

Each jet is stored as a list of up to 200 reconstructed constituents. For each constituent, the dataset provides its four-momentum,
\[
p_i^\mu=(E_i,p_{x,i},p_{y,i},p_{z,i}),
\qquad i=1,\dots,N_{\text{CONST}}.
\]
Jets with fewer than 200 constituents are zero-padded, while jets with more constituents are truncated by construction of the benchmark. In the released dataset, the constituents are ordered by decreasing transverse momentum.

For the present study, we retain only the first
\[
N_{\text{CONST}}=48
\]
constituents. This choice keeps the input dimension manageable while preserving the dominant substructure information in the kinematic regime under consideration. Just as importantly, it provides every method with the same fixed-size input, which is necessary for a controlled comparison.

To quantify the information retained by this choice, we evaluated the cumulative transverse-momentum coverage of the leading constituents in the benchmark dataset. The leading 48 constituents retain on average 98.9\% of the total constituent transverse momentum, compared with 95.8\% for 32 constituents and 99.8\% for 64 constituents. We therefore adopt $N_{\rm CONST}=48$ as a practical compromise between information retention and computational cost. The fact that $48 = 3\times 2^4$ is compatible with the four-layer binary MERA hierarchy used in this work, but this was not the primary motivation for the choice.

Rather than using raw Cartesian momentum components directly, we express each jet in a jet-centred coordinate system. The four-momentum reconstructed from the retained constituents is
\[
p^\mu_{\text{jet}}=\sum_{i=1}^{N_{\text{CONST}}}p_i^\mu,
\]
from which the jet axis $(p_{T,\text{jet}},\eta_{\text{jet}},\phi_{\text{jet}})$ is obtained. For each constituent, we compute the transverse momentum
\[
p_T^i=\sqrt{p_{x,i}^2+p_{y,i}^2},
\]
the azimuthal angle
\[
\phi_i=\mathrm{atan2}(p_{y,i},p_{x,i}),
\]
and the pseudorapidity
\[
\eta_i=\frac{1}{2}\ln\left(\frac{|\mathbf{p}_i|+p_{z,i}}{|\mathbf{p}_i|-p_{z,i}}\right),
\qquad
|\mathbf{p}_i|=\sqrt{p_{x,i}^2+p_{y,i}^2+p_{z,i}^2}.
\]
The constituent coordinates used as inputs are then
\[
\left(p_T^i,\Delta\eta_i,\Delta\phi_i\right),
\]
with
\[
\Delta\eta_i=\eta_i-\eta_{\text{jet}},
\qquad
\Delta\phi_i=\mathrm{wrap}\!\left(\phi_i-\phi_{\text{jet}}\right),
\]
where $\mathrm{wrap}$ maps the azimuthal difference to the interval $(-\pi,\pi]$.

This representation removes the trivial dependence on the global jet direction and focuses the model on the internal energy flow within the jet. In other words, the models see how momentum is distributed \emph{inside} the jet rather than where the jet happens to point in the detector. The final input is obtained by concatenating the constituent triplets into a vector of dimension
\[
3\times N_{\text{CONST}}=144.
\]
The same representation is used for all neural and classical baselines.

\subsection{Locality-Preserving Ordering and Standardisation}

The original benchmark orders constituents by decreasing $p_T$. This is a sensible default, but it is not ideal for a multiscale architecture whose local tensor operations act on neighbouring sites. For a MERA-type model, the notion of "neighbouring" constituents matters: one would like adjacent sites to correspond, as far as possible, to constituents that are also close in the jet.

For this reason, we reorder the active constituents using a simple geometry-based procedure in the $(\Delta\eta,\Delta\phi)$ plane. Constituents with
\[
p_T \le 10^{-6}\,\mathrm{GeV}
\]
are treated as inactive padding entries. This is not a physical transverse-momentum cut, but only a numerical zero threshold used to identify padded constituents in the native dataset units. Among the active constituents, we begin with the highest-$p_T$ constituent and then append the nearest unvisited neighbour according to the angular distance
\begin{equation}
\Delta R_{ij}^2 =
(\Delta\eta_i-\Delta\eta_j)^2
+
(\Delta\phi_i-\Delta\phi_j)^2 .
\end{equation}
Inactive padded entries are placed at the end of the sequence. This does not change the jet itself, but it produces a fixed ordering that better preserves local geometric correlations. The same reordered inputs are used for every model in the benchmark, so any performance differences cannot be attributed to giving MERA privileged information.

After ordering, each input feature is standardised using only the background portion of the training set. If $x_j$ denotes a given input component, we define
\begin{equation}
\tilde{x}_j=\frac{x_j-\mu_j}{\sigma_j+10^{-6}},
\end{equation}
where $\mu_j$ and $\sigma_j$ are the mean and standard deviation computed from background training jets alone. The additive term $10^{-6}$ is a small numerical stabiliser: some components can have vanishing or extremely small variance, especially for padded entries, and the offset prevents division by zero or by an anomalously small denominator. The same transformation is then applied to validation and test data. This prevents information leakage from the signal sample and ensures that every method sees identically normalised inputs. To improve numerical stability, we additionally clip the standardised features to the range
\begin{equation}
\tilde{x}_j \leftarrow \mathrm{clip}(\tilde{x}_j,-z_{\max},z_{\max}),
\qquad
z_{\max}=5.
\end{equation}
We checked explicitly how often this clipping threshold is reached before clipping is applied. Only approximately $0.12\%$ of feature entries in QCD jets and $0.086\%$ in top jets lie outside the interval $[-5,5]$. Clipping therefore affects a negligible fraction of entries in both samples and does not preferentially remove extreme values from the top-jet sample.

We also decomposed the per-jet mean-squared reconstruction loss into its $p_T$, $\Delta\eta$, and $\Delta\phi$ contributions after standardisation. Averaged over three independent seeds at $B=8$, the $p_T$ component accounts for approximately $60\%$ of the dense-autoencoder loss on QCD jets and $72\%$ on top jets. For the MERA-inspired model, the corresponding fractions are approximately $52\%$ and $61\%$. Thus, $p_T$ provides the largest single contribution for both architectures and both samples, while the two angular variables together account for the remainder. These fractions refer to the reconstruction loss and do not alter the equal component-wise standardisation applied to the inputs.

\subsection{Benchmarking Protocol}

All methods are compared in the same unsupervised setting. Neural models are trained on background jets only, and the classical baselines are likewise fitted only to the background training sample. Signal jets are withheld until validation and testing. This benchmarking protocol is designed to answer a specific question: how well can each method learn the structure of ordinary QCD jets, and how strongly does a top jet deviate from that learned background description?

For the neural models studied here, anomaly scores are defined by the per-jet reconstruction loss. A jet that is well represented by the learned background manifold receives a small score, while a jet that cannot be reconstructed accurately receives a larger score and is interpreted as more anomalous. For the classical baselines, we use the natural score associated with each method: the reconstruction residual for PCA, the Mahalanobis distance for the Gaussian model, and the isolation-depth-based score for Isolation Forest.

Because the dataset, preprocessing, train--validation--test splits, and scoring protocol are fixed across all models, the benchmark isolates the effect of representational structure as cleanly as possible. This is the main reason for organising the study in this way.

\section{Model Implementations and Reference Baselines}
\label{sec:models}

This section specifies the concrete models used in the numerical study. All methods act on the common representation defined in \Sec{sec:data}, namely a standardised vector
\begin{equation}
x\in\mathbb{R}^{144},
\end{equation}
obtained from $48$ ordered jet constituents with three features each. For the capacity-controlled comparison, MERA, the dense autoencoder, and PCA are studied at effective latent dimension $B\in\{8,16,32\}$. The Gaussian and Isolation Forest baselines do not contain an explicit bottleneck, but they are trained on the same background-only sample and evaluated on the same validation and test sets. Details of optimisation and model selection are summarised at the beginning of \Sec{sec:results}; here, the focus is on the structure of the models themselves.

\subsection{MERA-Inspired Neural Implementation}
\label{sec:mera_arch}

The canonical MERA construction and its motivation for anomaly detection were discussed in \Sec{sec:mera_construction}. We now define the modified, finite MERA-inspired neural architecture used in the benchmark. In particular, the implementation uses within-pair rather than staggered cross-boundary mixing, nonlinear activations between hierarchical levels, an untied decoder, and soft structural constraints. The input vector is reshaped into a sequence of site tensors,
\begin{equation}
x \mapsto X^{(0)}=\{s_i^{(0)}\}_{i=1}^{48},
\qquad
s_i^{(0)}\in\mathbb{R}^{3},
\end{equation}
so that each site corresponds to one ordered constituent described by $(p_T,\Delta\eta,\Delta\phi)$. We then apply four binary coarse-graining stages, giving
\begin{equation}
n_0=48,\qquad n_1=24,\qquad n_2=12,\qquad n_3=6,\qquad n_4=3,
\end{equation}
where $n_\ell$ denotes the number of active sites at level $\ell$. In the implementation used here, the site dimension is kept fixed at $\chi=3$ at every scale. This keeps the local tensors small and makes the trainable complexity grow mainly with the number of sites rather than with an increasing bond dimension.

At each layer $\ell=0,\dots,3$, neighbouring sites are paired according to the geometry-preserving ordering introduced in \Sec{sec:data}. For the $j$-th pair, we form the local vector
\begin{equation}
p_j^{(\ell)}=
\begin{pmatrix}
s_{2j-1}^{(\ell)}\\
s_{2j}^{(\ell)}
\end{pmatrix}
\in\mathbb{R}^{6}.
\end{equation}
A learned local mixing map, playing the role of a disentangler in the MERA-inspired construction,
\begin{equation}
U_j^{(\ell)}\in\mathbb{R}^{6\times 6}
\end{equation}
first mixes the two neighbouring sites, and a learned coarse-graining map, playing the role of an isometry,
\begin{equation}
W_j^{(\ell)}\in\mathbb{R}^{3\times 6}
\end{equation}
then projects the result to the next scale:
\begin{equation}
\tilde p_j^{(\ell)} = U_j^{(\ell)} p_j^{(\ell)},
\qquad
s_j^{(\ell+1)} = \sigma\!\left(W_j^{(\ell)} \tilde p_j^{(\ell)}\right),
\end{equation}
where $\sigma$ denotes a leaky-ReLU activation with slope parameter $\alpha=0.1$ at the intermediate hierarchical levels and the identity at the final encoder level. Thus, the nonlinearity is applied after each intermediate coarse-graining operation $W_j^{(\ell)}$. Unlike canonical MERA, the implemented layer is nonlinear and its $U_j^{(\ell)}$ and $W_j^{(\ell)}$ maps act successively on the same pair $(s_{2j-1}^{(\ell)},s_{2j}^{(\ell)})$. The tensors are not shared across positions or layers. This is a deliberate choice: the ordered jet is finite, not translationally invariant, and the upper part of the hierarchy is irregular because $48\neq 2^L$.

After the final coarse-graining stage, the remaining three sites are concatenated into a top tensor
\begin{equation}
h\in\mathbb{R}^{9}.
\end{equation}
A linear map then produces the latent representation,
\begin{equation}
z = P_{\mathrm{enc}} h,
\qquad
P_{\mathrm{enc}}\in\mathbb{R}^{B\times 9},
\end{equation}
with $B\in\{8,16,32\}$. For $B=16$ and $32$, this final map expands the $9$-dimensional top tensor, so the main compression is already carried by the hierarchical coarse-graining rather than by this last linear layer alone. The decoder begins with an untied linear expansion
\begin{equation}
\hat h = P_{\mathrm{dec}} z,
\qquad
P_{\mathrm{dec}}\in\mathbb{R}^{9\times B},
\end{equation}
and then reconstructs finer scales through learned expansion maps
\begin{equation}
\bar W_j^{(\ell)}\in\mathbb{R}^{6\times 3}
\end{equation}
and local mixing maps
\begin{equation}
\bar U_j^{(\ell)}\in\mathbb{R}^{6\times 6},
\end{equation}
according to
\begin{equation}
\tilde q_j^{(\ell)} = \bar W_j^{(\ell)} s_j^{(\ell+1)},
\qquad
\hat q_j^{(\ell)} = \bar U_j^{(\ell)} \tilde q_j^{(\ell)},
\end{equation}
followed by reshaping $\hat q_j^{(\ell)}$ into two daughter sites at the next finer level. The final output layer is linear, so the model reconstructs
\begin{equation}
\hat x\in\mathbb{R}^{144}
\end{equation}
in the same feature space in which it was trained.

The network is trained by minimising the per-jet mean-squared reconstruction loss
\begin{equation}
\mathcal{L}_{\mathrm{rec}}(x)=\frac{1}{144}\|x-\hat x\|_2^2,
\end{equation}
augmented by soft orthogonality penalties on the tensors that play the role of disentanglers and isometries,
\begin{equation}
\mathcal{L}
=
\mathcal{L}_{\mathrm{rec}}
+
\lambda_U\sum_{\ell,j}\bigl\|{U_j^{(\ell)}}^\top U_j^{(\ell)}-I_6\bigr\|_F^2
+
\lambda_W\sum_{\ell,j}\bigl\|W_j^{(\ell)}{W_j^{(\ell)}}^\top-I_3\bigr\|_F^2.
\end{equation}
In the nominal benchmark, we use \(\lambda_U=\lambda_W=10^{-3}\). Rather than enforcing exact constraints on orthogonal or Stiefel manifolds, we use these penalties to keep the learned maps close to the intended MERA structure while retaining the flexibility of end-to-end optimisation~\cite{Vidal:2007er,Vidal:2008mera,Orus:2014intro,Stoudenmire:2018multiscale,Reyes:2021multiscale,Batselier:2021meracle}. The decoder is kept untied for a related reason: after coarse-graining, exact inversion is impossible, so a learned approximate inverse is more suitable than a rigid transpose construction. Once trained on background jets only, the anomaly score is taken to be the reconstruction error,
\begin{equation}
S_{\mathrm{MERA}}(x)=\mathcal{L}_{\mathrm{rec}}(x).
\end{equation}

\begin{figure}[H]
    \centering
    \includegraphics[width=0.9\textwidth]{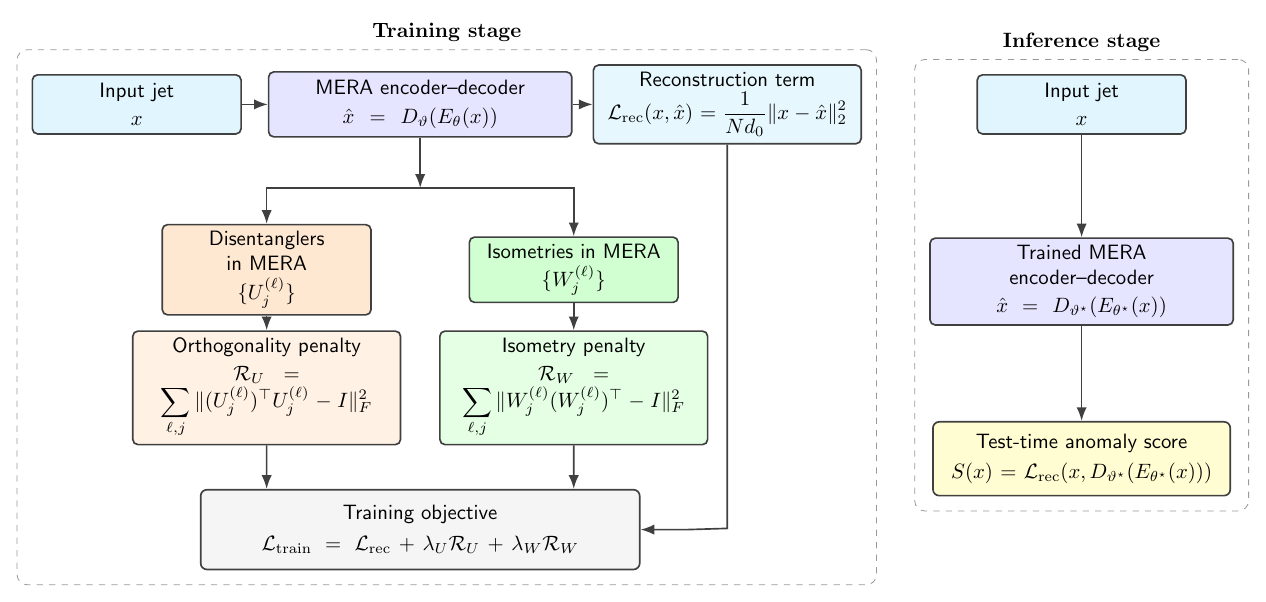}
    \caption{Schematic relationship between the training objective and inference-time anomaly score for the implemented MERA-inspired network. During training, the reconstruction loss is supplemented by soft orthogonality and isometry penalties on the local mixing and coarse-graining maps. At inference time, only the per-jet reconstruction error is used as the anomaly score.}
    \label{fig:mera_training_objective}
\end{figure}

The relationship between the concrete penalised training objective and the inference-time anomaly score is illustrated in \Fig{fig:mera_training_objective}. During training, the reconstruction and regularisation terms act on the same encoder--decoder parameters, whereas at inference time only the reconstruction error is used as the anomaly score.

\subsection{Dense Autoencoder Baseline}
\label{sec:ae_baseline}

Fully connected autoencoders provide the standard nonlinear reconstruction baseline for unsupervised anomaly detection and have been widely used in collider applications~\cite{Hinton:2006autoencoder,Heimel:2019qcd,Farina:2020deepae,Collins:2021weakunsup,Barron:2021suep,Blance:2019adversarialae,Atkinson:2022ircgraphae,Finke:2021autoencoders}. Their role in the present study is clear: unlike MERA, a dense autoencoder can mix all input features globally, but it carries no explicit notion of scale or locality. Any hierarchical organisation in the jet therefore has to be discovered implicitly from the training data rather than being built into the architecture.

Our reference autoencoder takes the same $144$-dimensional input and uses a symmetric fully connected encoder--decoder of the form
\begin{equation}
144 \rightarrow 48 \rightarrow 24 \rightarrow B \rightarrow 24 \rightarrow 48 \rightarrow 144,
\end{equation}
with $B\in\{8,16,32\}$. Writing the encoder and decoder as $f_{\mathrm{enc}}$ and $f_{\mathrm{dec}}$, the model computes
\begin{equation}
z=f_{\mathrm{enc}}(x)\in\mathbb{R}^{B},
\qquad
\hat x=f_{\mathrm{dec}}(z)\in\mathbb{R}^{144}.
\end{equation}
Leaky-ReLU activations with slope $\alpha=0.1$ are used in the hidden layers, while the bottleneck and output layers are kept linear. The architecture is intentionally modest: it is expressive enough to model global nonlinear correlations, but still simple enough that differences with respect to MERA can be interpreted as differences in inductive bias rather than brute-force network size.

The autoencoder is trained on background jets only with the same reconstruction objective used for MERA,
\begin{equation}
\mathcal{L}_{\mathrm{rec}}(x)=\frac{1}{144}\|x-\hat x\|_2^2,
\end{equation}
and the anomaly score is defined by
\begin{equation}
S_{\mathrm{AE}}(x)=\mathcal{L}_{\mathrm{rec}}(x).
\end{equation}
Using the same input representation, the same loss, and the same latent sizes makes this model a direct test of whether explicit multiscale structure improves background compression over a generic dense map.

\subsection{Classical Reference Baselines}
\label{sec:classical_baselines}

To complement the neural comparison, we also include three standard unsupervised baselines that embody different notions of normality: a linear low-rank model (PCA), a single-component density model (Gaussian/Mahalanobis), and a tree-based isolation method (Isolation Forest). All are fitted on the same standardised background sample used for the neural architectures.

\paragraph{Principal Component Analysis.}
PCA is the canonical linear compression baseline~\cite{Pearson:1901pca,Hotelling:1933pca}. Let $\hat\mu$ denote the empirical background mean and let
\begin{equation}
V_B=\bigl(v_1,\dots,v_B\bigr)\in\mathbb{R}^{144\times B}
\end{equation}
contain the $B$ leading eigenvectors of the empirical covariance matrix. The rank-$B$ reconstruction is
\begin{equation}
\hat x_{\mathrm{PCA}}=\hat\mu + V_BV_B^\top(x-\hat\mu),
\end{equation}
and the anomaly score is the squared residual
\begin{equation}
S_{\mathrm{PCA}}(x)=\|x-\hat x_{\mathrm{PCA}}\|_2^2
=
\bigl\|(I-V_BV_B^\top)(x-\hat\mu)\bigr\|_2^2.
\end{equation}
PCA is especially informative in the present context because each MERA isometry performs a local linear truncation. It therefore provides the cleanest global linear reference against which one can assess the value of multiscale nonlinear structure.

\paragraph{Gaussian Background Model.}
As a minimal density-estimation baseline, we fit a multivariate Gaussian distribution to the background sample using the empirical mean $\hat\mu$ and a ridge-regularised covariance matrix $\hat\Sigma+\lambda_{\mathrm{G}}I$ with $\lambda_{\mathrm{G}}=10^{-3}$. Under this model, the negative log-likelihood differs from the squared Mahalanobis distance only by additive terms that are common across events, so we use the corresponding regularised Mahalanobis score~\cite{Mahalanobis:1936distance}
\begin{equation}
S_{\mathrm{G}}(x)=(x-\hat\mu)^\top \bigl(\hat\Sigma+\lambda_{\mathrm{G}}I\bigr)^{-1}(x-\hat\mu).
\end{equation}
This baseline asks whether the standardised background can already be described adequately by a single ellipsoidal distribution in feature space. If the background manifold is curved, multimodal, or strongly non-Gaussian, this model should underperform, which helps to interpret the gains from more structured methods.

\paragraph{Isolation Forest.}
Isolation Forest provides a conceptually different unsupervised baseline based on recursive random partitioning rather than reconstruction or density estimation~\cite{Liu:2008iforest}. We use an ensemble of $T=100$ isolation trees trained on background events. For an event $x$, each tree returns a path length $h_t(x)$, and the average path length
\begin{equation}
\bar h(x)=\frac{1}{T}\sum_{t=1}^{T}h_t(x)
\end{equation}
is converted into the standard isolation score
\begin{equation}
S_{\mathrm{IF}}(x)=2^{-\bar h(x)/c(n)},
\qquad
c(n)=2H_{n-1}-\frac{2(n-1)}{n},
\end{equation}
where $n$ is the subsample size used to construct a tree and $H_k$ denotes the $k$-th harmonic number. Jets that are isolated after fewer random splits receive larger anomaly scores. Any monotonic equivalent of this standard score would lead to the same ROC ranking, but the expression above makes the statistical interpretation transparent.

\section{Results and Discussion}
\label{sec:results}
\label{sec:training}

For both MERA and the dense autoencoder, hyperparameters were selected separately for each bottleneck dimension \(B\) using the validation sample. The selected configurations were then fixed and evaluated over independent random seeds disjoint from those used for model selection. All methods are trained on the QCD background sample only, with signal events used exclusively for validation and final testing. For the neural architectures, optimisation is carried out with Adam~\cite{Kingma:2014vow} using the reconstruction losses defined in \Sec{sec:models}, while the classical reference models are fitted with their standard unsupervised objectives. For stochastic models, the final performance is reported as a mean over repeated runs with independent random seeds.

The main question of this section is no longer just which curve sits highest in one benchmark table. We instead separate two issues. First, we establish the nominal benchmark comparison against the dense autoencoder and the classical baselines. Second, we test the locality argument directly through a training-free local-compressibility diagnostic and then ask whether the disentanglers improve on the corresponding identity-disentangler limit of the same hierarchy.

\subsection{Nominal benchmark}

\begin{figure}[H]
\centering

\begin{subfigure}{\textwidth}
\centering
\includegraphics[width=0.32\textwidth]{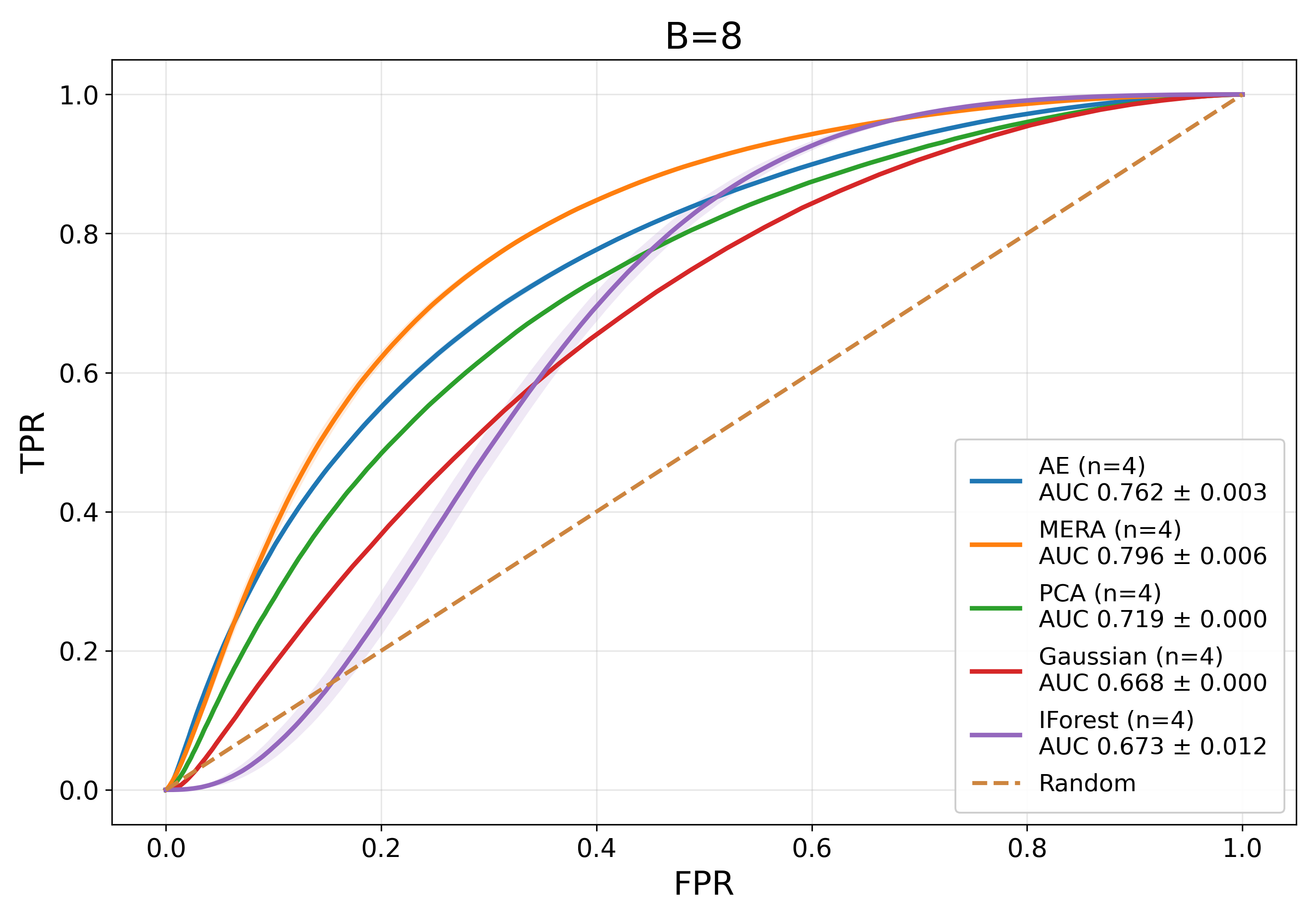}
\includegraphics[width=0.32\textwidth]{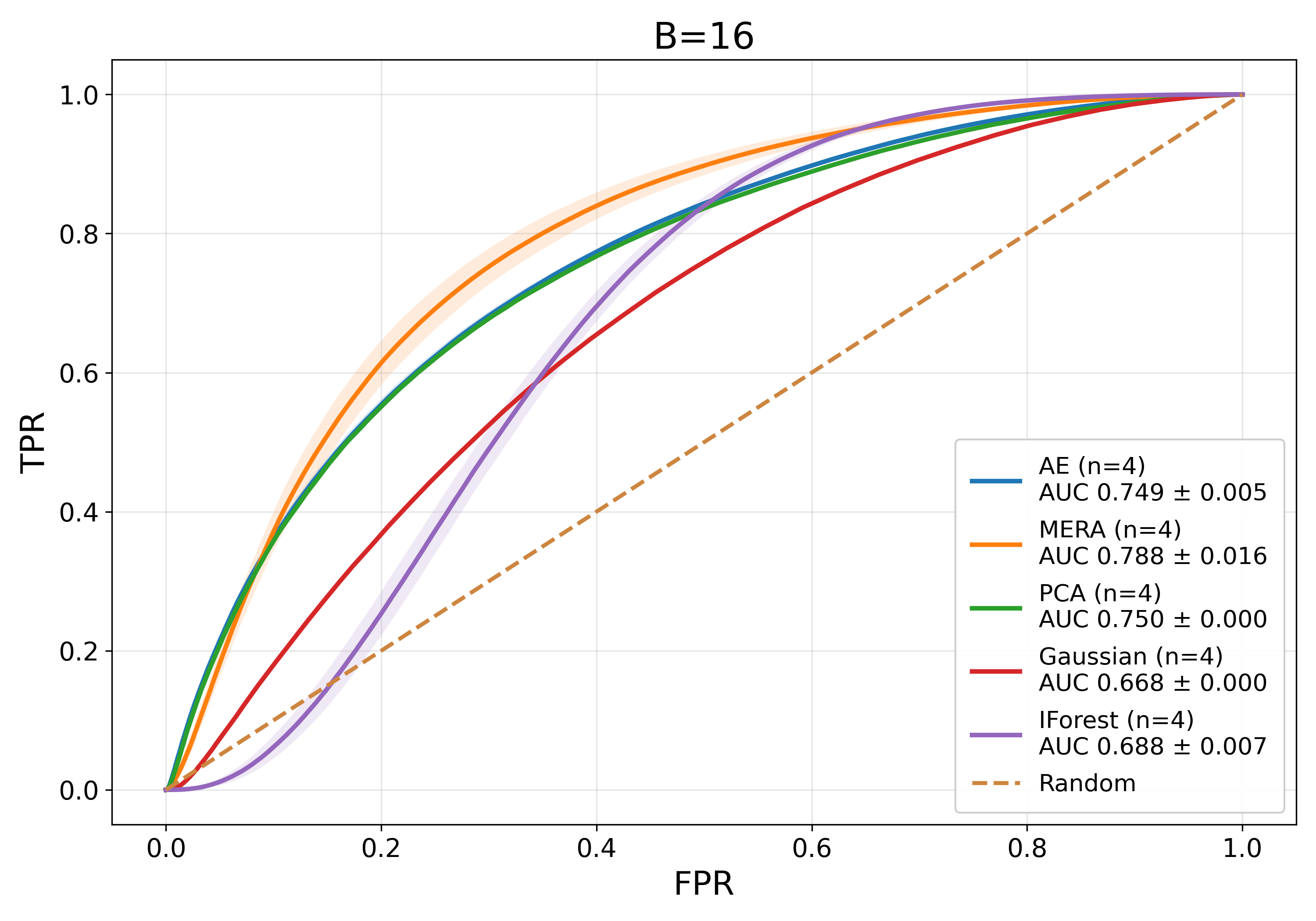}
\includegraphics[width=0.32\textwidth]{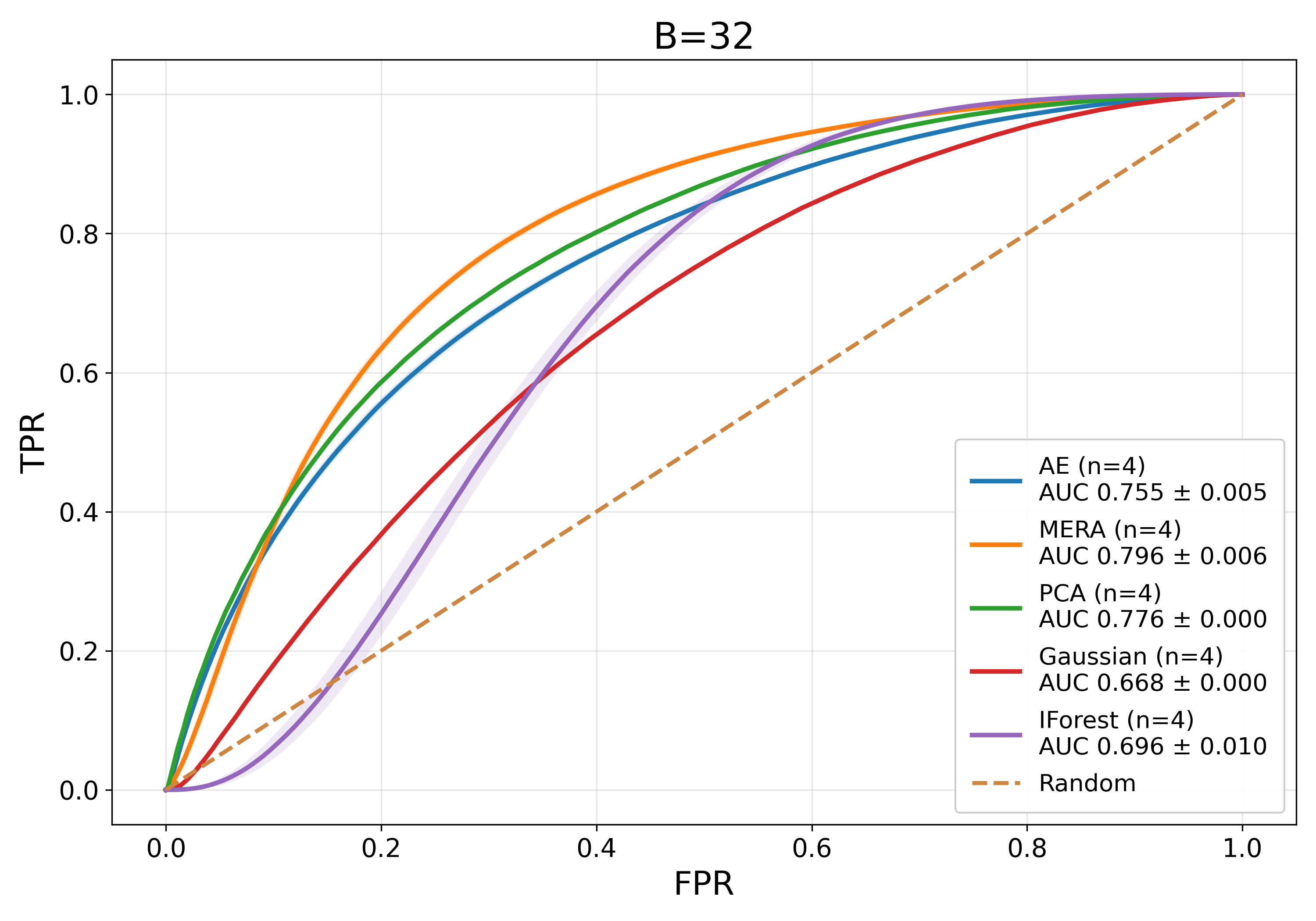}
\caption{ROC curves at latent dimensions $B=8,16,32$ on a linear false-positive-rate scale.}
\end{subfigure}

\vspace{0.4cm}

\begin{subfigure}{\textwidth}
\centering
\includegraphics[width=0.32\textwidth]{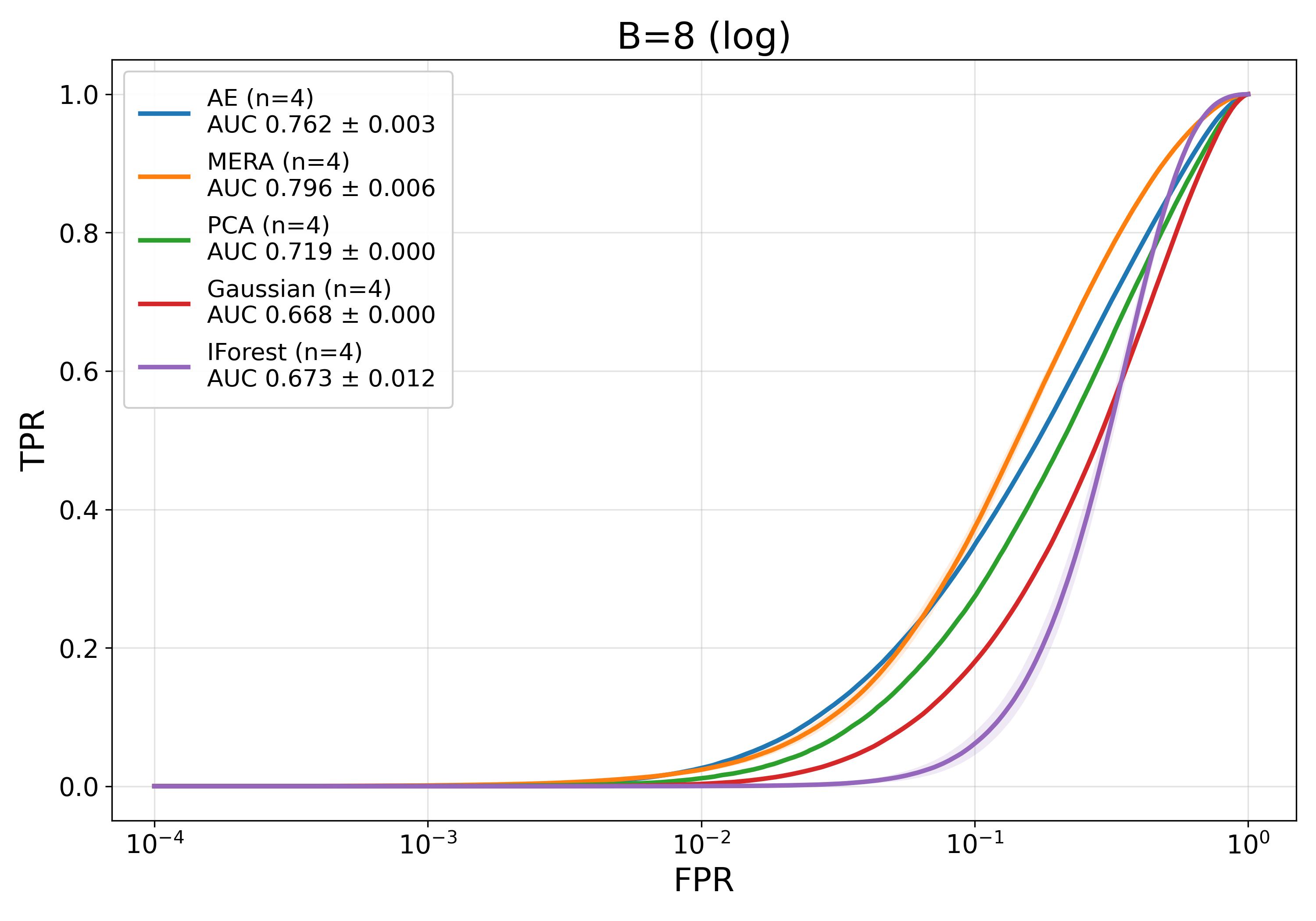}
\includegraphics[width=0.32\textwidth]{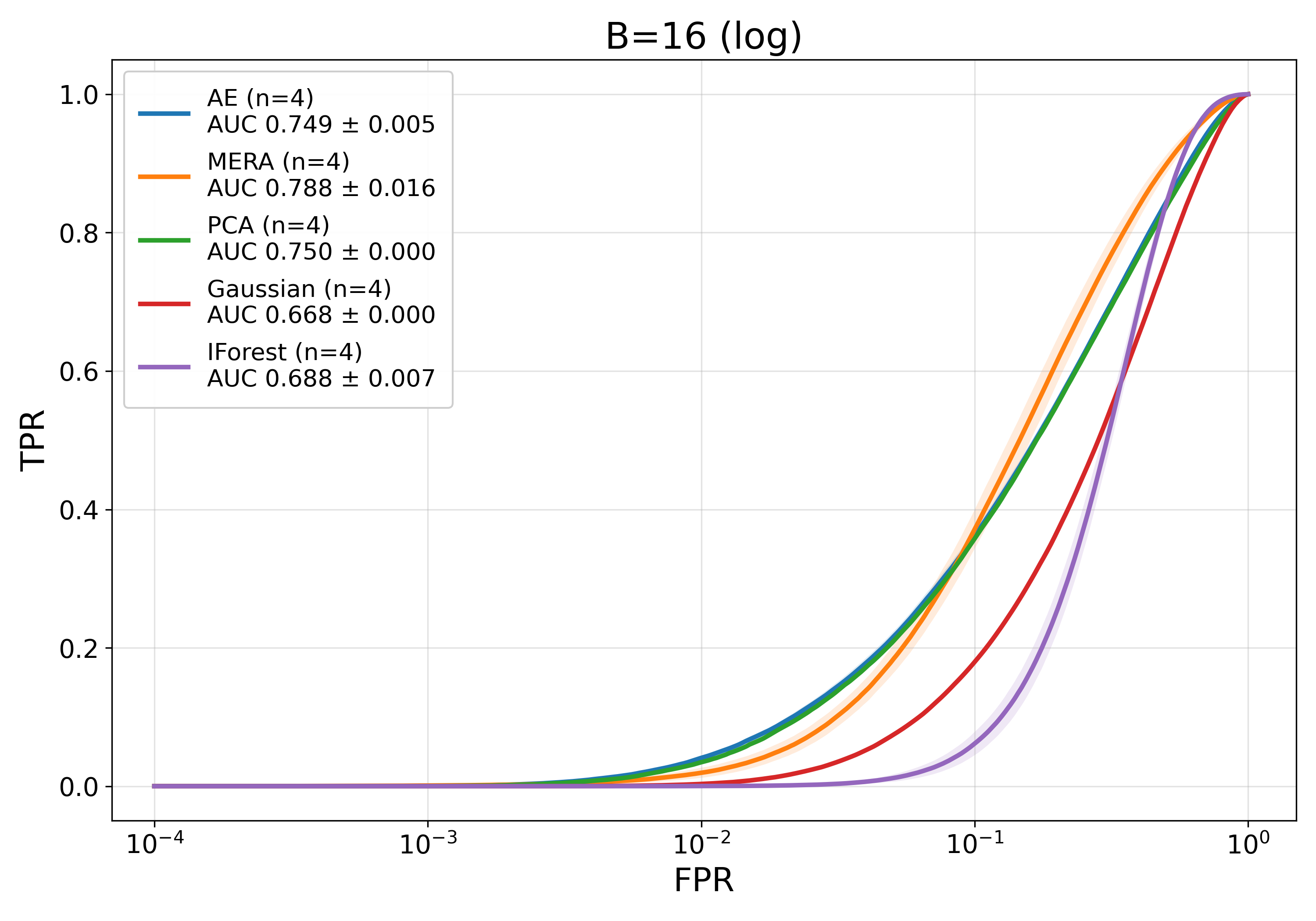}
\includegraphics[width=0.32\textwidth]{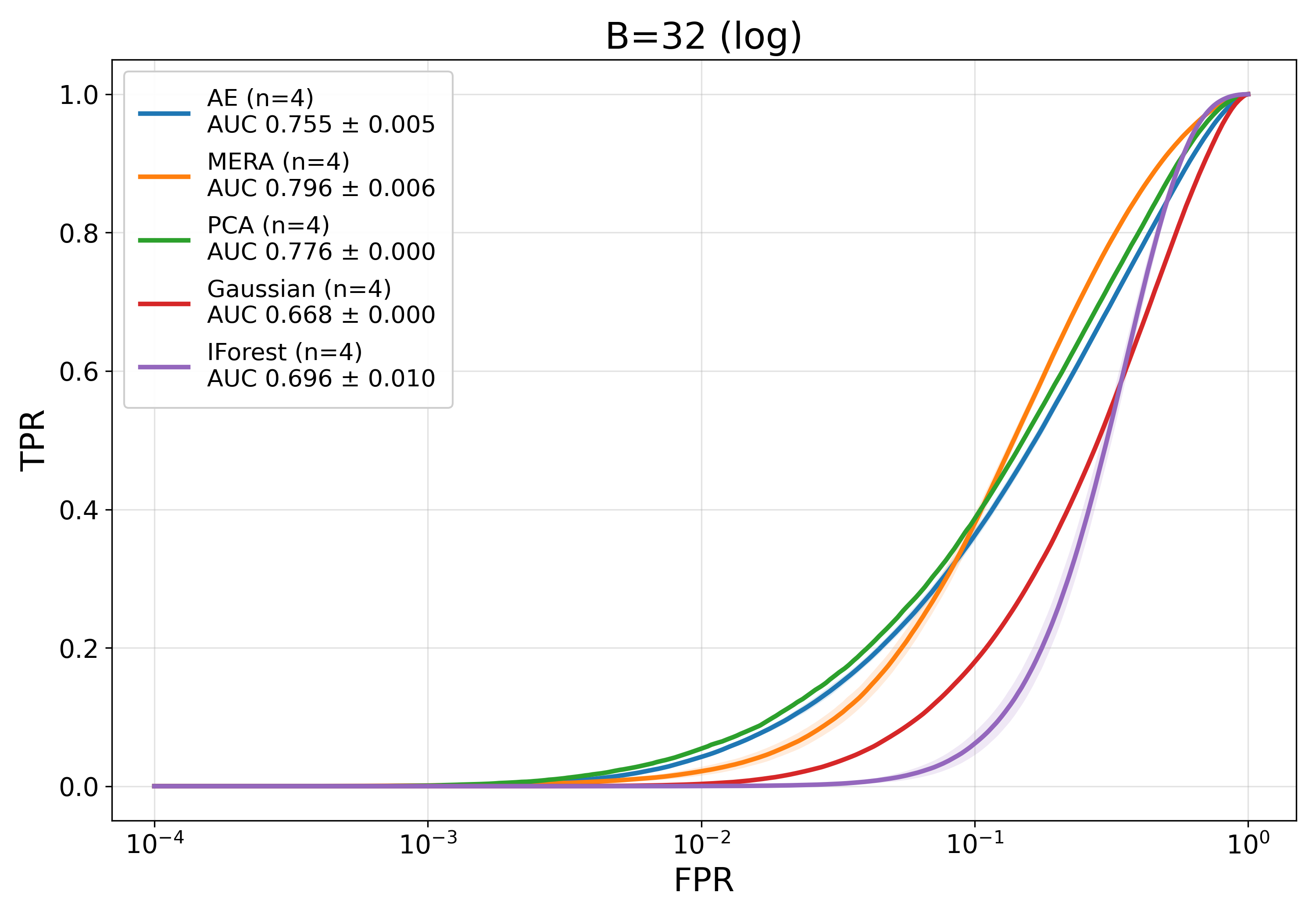}
\caption{The same ROC curves in the low-false-positive-rate regime, shown on a logarithmic false-positive-rate scale.}
\end{subfigure}

\caption{
ROC curves for the nominal benchmark at latent dimensions $B=8,16,32$. Solid lines denote the mean over random seeds where applicable, and shaded bands indicate the corresponding $\pm1\sigma$ variation.
}
\label{fig:roc_curves}
\end{figure}

Figure~\ref{fig:roc_curves} and Table~\ref{tab:results} summarise the nominal benchmark for the locality-preserving representation introduced in \Sec{sec:data}. Within this specific scan, MERA attains the largest test AUC at all three latent dimensions, reaching $0.7959$, $0.7885$, and $0.7963$ at $B=8,16,32$, respectively. The corresponding dense-autoencoder values are $0.7616$, $0.7487$, and $0.7550$, while PCA becomes competitive only once the retained dimension is large enough to capture a sizeable fraction of the global background variance.

The comparison to the dense autoencoder is the most important one in this subsection, because the dense AE is the standard nonlinear reconstruction baseline and carries substantially more trainable freedom than the implemented MERA-inspired model. The MERA-inspired model attains a higher mean AUC at every latent dimension in this benchmark: the gain is about $0.034$ at $B=8$, $0.040$ at $B=16$, and $0.041$ at $B=32$. At the same time, the implemented models use only about one third of the trainable parameters of the corresponding dense autoencoders: roughly $5.0$--$5.5\times10^3$ parameters, compared with $1.68$--$1.80\times10^4$ for AE. In other words, these numerical results show that, for the locality-preserving representation used here, the implemented MERA-inspired multiscale construction can outperform the standard dense autoencoder while doing so with a far smaller parameter budget. This empirical statement concerns the neural architecture defined in \Sec{sec:mera_arch}, not canonical MERA in general.

That point is important for the logic of the paper. If the MERA model had only matched the AE after introducing extra structure, the result would mainly be architectural curiosity. Instead, Table~\ref{tab:results} shows a simultaneously stronger and more economical model in the nominal scan. This is exactly the kind of outcome that motivates asking which part of the construction is responsible: whether the gain is tied to the locality-aligned representation, to hierarchical coarse-graining more generally, or to the disentangling layers specific to MERA.

Table~\ref{tab:results} shows the same comparison numerically. The key observation is therefore quite concrete: in the benchmark setting used throughout this paper, the implemented MERA-inspired architecture attains a higher mean AUC than the dense autoencoder with only about one-third of the dense model's trainable parameters. This makes the nominal benchmark a positive result for the specific neural construction studied here and provides the starting point for the more targeted tests below.

\begin{table}[htbp]
\centering
\small
\caption{Test-set ROC-AUC values for all models at latent dimensions $B=8,16,32$, where AUC denotes the area under the receiver operating characteristic (ROC) curve and CI denotes the confidence interval quoted for repeated runs. For the neural architectures, the corresponding number of trainable parameters is also listed. For PCA and the Gaussian baseline, the reported AUC standard deviations are exactly zero because these methods are deterministic once the dataset and preprocessing are fixed.}
\setlength{\tabcolsep}{4pt}
\resizebox{\textwidth}{!}{
\begin{tabular}{c l l c c c c c}
$B$ & Model & Architecture & Params & Test AUC & AUC(std) & CI low & CI high \\
\midrule
8  & MERA     & (L=4, latent=8, $lr=3\times10^{-4}$)          & 5021  & 0.7959 & 0.0065 & 0.7855 & 0.8063 \\
8  & AE       & (h1=48, h2=24, bn=8, $lr=3\times10^{-4}$)     & 16808 & 0.7616 & 0.0038 & 0.7554 & 0.7677 \\
8  & PCA      & (ncomp=8)                                      & --    & 0.7192 & 0.0000 & 0.7192 & 0.7192 \\
8  & Gaussian & Mahalanobis (ridge-reg.\ cov)                 & --    & 0.6679 & 0.0000 & 0.6679 & 0.6679 \\
8  & IForest  & (ntrees=100)                                   & --    & 0.6726 & 0.0142 & 0.6500 & 0.6952 \\
\hline
\midrule
16 & MERA     & (L=4, latent=16, $lr=2\times10^{-4}$)         & 5173  & 0.7885 & 0.0190 & 0.7582 & 0.8189 \\
16 & AE       & (h1=48, h2=24, bn=16, $lr=3\times10^{-4}$)   & 17200 & 0.7487 & 0.0058 & 0.7394 & 0.7579 \\
16 & PCA      & (ncomp=16)                                     & --    & 0.7502 & 0.0000 & 0.7502 & 0.7502 \\
16 & Gaussian & Mahalanobis (ridge-reg.\ cov)                 & --    & 0.6679 & 0.0000 & 0.6679 & 0.6679 \\
16 & IForest  & (ntrees=100)                                   & --    & 0.6880 & 0.0082 & 0.6749 & 0.7011 \\
\hline
\midrule
32 & MERA     & (L=4, latent=32, $lr=1.5\times10^{-4}$)       & 5477  & 0.7963 & 0.0074 & 0.7845 & 0.8081 \\
32 & AE       & (h1=48, h2=24, bn=32, $lr=2\times10^{-4}$)   & 17984 & 0.7550 & 0.0053 & 0.7466 & 0.7635 \\
32 & PCA      & (ncomp=32)                                     & --    & 0.7764 & 0.0000 & 0.7764 & 0.7764 \\
32 & Gaussian & Mahalanobis (ridge-reg.\ cov)                 & --    & 0.6679 & 0.0000 & 0.6679 & 0.6679 \\
32 & IForest  & (ntrees=100)                                   & --    & 0.6957 & 0.0119 & 0.6767 & 0.7147 \\
\hline
\end{tabular}
}
\label{tab:results}
\end{table}

\subsection{Locality and local compressibility}

The most direct way to test the locality argument of \Sec{sec:mera_construction} is to ask whether the ordered input is actually more compressible at the scale on which the local tensors of the implemented MERA-inspired architecture act. We therefore performed a training-free diagnostic on background jets for the three constituent orderings considered earlier: the locality-preserving chain used in the main benchmark, a simple $p_T$ ordering, and random permutations. For each ordering, we measured the angular separation between adjacent sites and the retained variance fraction of the optimal three-dimensional approximation to neighbouring two-site blocks. The latter quantity is the local linear analogue of the compression step performed by the first layers of the implemented model.

Table~\ref{tab:local_compressibility} summarises the two quantities that matter most for this question. The mean adjacent-site separation $\langle \Delta R \rangle$ measures how geometrically local neighbouring entries in the one-dimensional MERA chain actually are. The retained-variance fractions then quantify how compressible those neighbouring two-site blocks are under the optimal three-dimensional approximation, providing a linear proxy for the first local compression step of the hierarchy.

As shown in Table~\ref{tab:local_compressibility}, the locality-preserving ordering reduces the mean adjacent-site separation to $\langle\Delta R\rangle=0.131$, compared with $0.361$ for $p_T$ ordering and $0.397\pm0.004$ for random permutations. More importantly, the same ordering yields the highest retained variance in the first two compression layers: $0.869$ and $0.920$, compared with $0.857$ and $0.890$ for $p_T$ ordering, and only $0.721$ and $0.830$ for the random control. The effect is strongest precisely where the MERA architecture is most local, namely in the first layers that combine neighbouring constituents before any coarse-graining has taken place.

This is also the physics reason to expect the implemented MERA-inspired architecture to benefit from a locality-preserving input representation. Jets are produced by a branching cascade, so short-range angular correlations are not random noise: they encode the local organisation of prongs, collinear radiation, and softer emissions around them. The implemented model is designed to process such a structure hierarchically. Its local maps first rearrange correlations inside neighbouring blocks and then coarse-grain those blocks scale by scale. If neighbouring slots in the input chain correspond to constituents that are genuinely close in angle, then the relevant short-distance correlations are presented where the local maps can exploit them most efficiently. By contrast, under $p_T$ ordering or random permutations, the first layers are forced to mix constituents that are often far apart inside the jet, so the local compression step is less well aligned with the physical organisation of the jet.

\begin{figure}[H]
\centering
\includegraphics[width=0.95\textwidth]{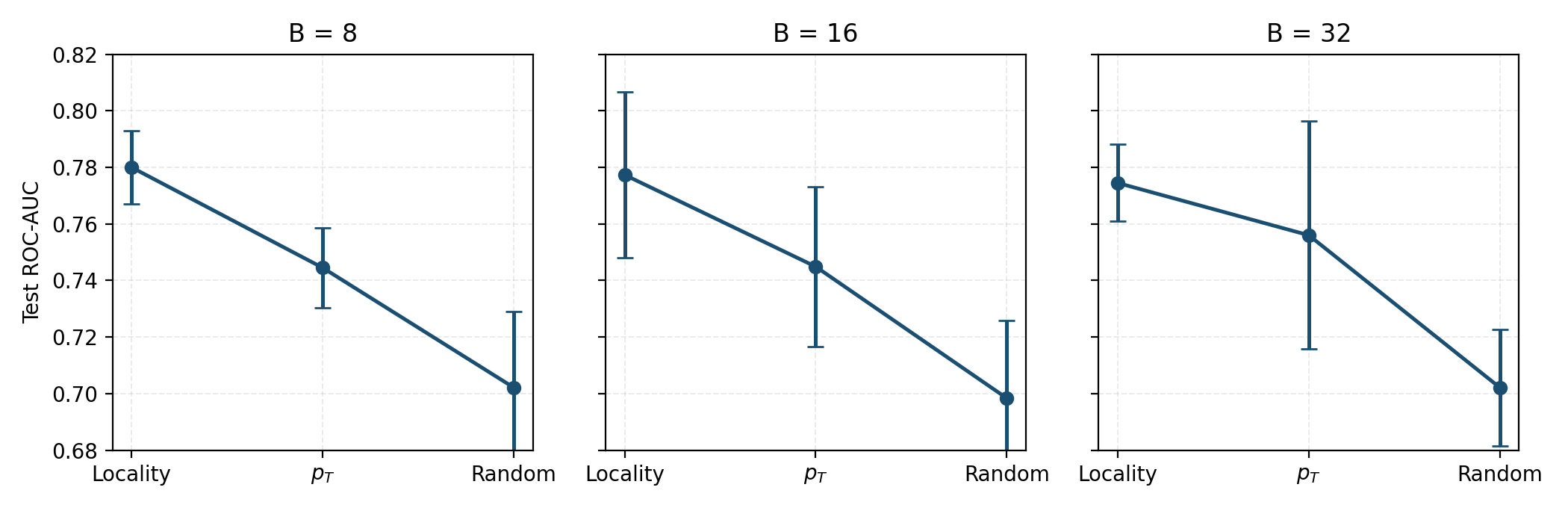}
\caption{MERA test ROC-AUC as a function of input ordering for $B=8,16,32$, aggregated over three independently tuned runs for each ordering. Error bars indicate the corresponding confidence intervals. The locality-preserving ordering gives the strongest MERA performance at every latent dimension, consistent with the geometric and compressibility diagnostics reported in Table~\ref{tab:local_compressibility}.}
\label{fig:mera_ordering_auc}
\end{figure}

The same pattern is reflected in the performance of the implemented MERA-inspired model. Figure~\ref{fig:mera_ordering_auc} summarises the completed ordering study for this model, based on three independently tuned runs for each ordering. The locality-preserving representation gives the highest mean test ROC-AUC at all three latent dimensions, with values of $0.780$, $0.777$, and $0.774$ at $B=8,16,32$, respectively. The corresponding $p_T$-ordered means are $0.745$, $0.745$, and $0.756$, while random permutations give $0.702$, $0.698$, and $0.702$. Thus, the ordering that makes adjacent input blocks most geometrically local and most compressible is also the ordering that gives the strongest performance for the implemented model on this benchmark. The separation is especially clear at $B=8$ and $B=16$, where the locality-ordered chain outperforms both alternatives by several percentage points in AUC.

To test whether this gain is a generic consequence of preprocessing rather than an architecture-specific effect, we repeated the same ordering study for the dense autoencoder and PCA. The dense autoencoder was retuned independently for each ordering and bottleneck using three selection replicates, with four evaluation seeds per replicate. Its mean AUC values for locality ordering are $0.747$, $0.745$, and $0.754$ at $B=8,16,32$, compared with $0.819$, $0.832$, and $0.831$ for $p_T$ ordering and $0.821$, $0.835$, and $0.834$ for random ordering. Thus, the dense model does not share the MERA improvement under locality ordering; it performs similarly for $p_T$ and random orderings and substantially worse for the locality-ordered representation. PCA likewise does not show a uniform locality benefit: its AUC changes from $0.834$, $0.836$, and $0.619$ for $p_T$ ordering to $0.719$, $0.750$, and $0.776$ for locality ordering. As expected for a global linear reconstruction model, PCA is invariant, up to numerical precision, under the fixed random permutations used here and therefore gives the same results for the $p_T$ and random representations. The contrast with MERA shows that the locality gain is not a shared improvement inherited by all models from the preprocessing. It arises specifically when the event-dependent local ordering is paired with an architecture whose first operations act on neighbouring sites.

Taken together, these results support the geometric premise of the construction. The locality-preserving chain is not just a convenient preprocessing choice: it aligns the ordered jet representation with the local compression steps of the MERA hierarchy and thereby places that model in the regime where its multiscale structure is most effective, while providing no corresponding general advantage to the dense or global-linear baselines.

\begin{table}[t]
\centering
\small
\caption{Geometric locality and two-site compressibility of the ordered constituent representation. The retained variance fractions are computed from the optimal three-dimensional approximation to neighbouring two-site blocks on the background sample.}
\label{tab:local_compressibility}
\begin{tabular*}{\textwidth}{@{\extracolsep{\fill}}lccc@{}}
\toprule
Ordering & Mean adjacent $\Delta R$ & Retained variance (layer 0) & Retained variance (layer 1) \\
\midrule
Locality & 0.131 & 0.869 & 0.920 \\
$p_T$    & 0.361 & 0.857 & 0.890 \\
Random   & $0.397 \pm 0.004$ & 0.721 & 0.830 \\
\bottomrule
\end{tabular*}
\end{table}

\subsection{Effect of the Disentanglers}

The direct architectural question is whether the trainable disentanglers in MERA improve anomaly detection beyond what is already achieved by the same hierarchy with purely local coarse-graining. To isolate that point as directly as possible, we compare the full MERA at the strongest-compression setting, $B=8$, with a constrained variant in which every disentangler is fixed to the identity map, denoted MERA($U=I$). This keeps the encoder--decoder hierarchy, latent map, and local isometries unchanged, but removes any learned local basis change before compression. Operationally, this choice makes the MERA mimic the corresponding TTN limit simply by replacing each disentangler with the identity while leaving the rest of the algorithm untouched. In the present implementation, this identity-disentangler limit is mathematically equivalent to the corresponding no-disentangler tree hierarchy, and it also ensures that the ablation introduces as few algorithmic changes as possible: the data representation, connectivity, training objective, optimisation procedure, and evaluation protocol are all kept fixed, so the comparison isolates the effect of the disentangling maps themselves.

The reason to focus on $B=8$ is also the reason to expect the clearest disentangler effect there. As discussed in \Sec{sec:mera_construction}, a TTN without disentanglers applies only the local projection $\hat q_{\mathrm{TTN}}=W^\top W q$, whereas MERA first rotates the local basis and only then truncates, $\hat q_{\mathrm{MERA}}=U^\top W^\top W U q$. The local truncation loss in \eqref{eq:local_truncation_loss} makes the role of $U$ explicit: the disentanglers are meant to reorganise short-range correlations so that the degrees of freedom retained by the isometries capture more of the informative variance before the next coarse-graining step. This mechanism should matter most precisely when the compression bottleneck is tight. At larger latent dimensions, the network can retain more information even without an optimal local rotation, whereas at $B=8$ the hierarchy is forced to be economical at every stage. This interpretation is consistent with the local-compressibility results of Table~\ref{tab:local_compressibility}, which already showed that the locality-preserving representation concentrates relevant structure into the neighbouring blocks on which the MERA tensors act.

The corresponding multi-seed comparison is shown in Table~\ref{tab:mera_disentangler_ablation}. At $B=8$, the full MERA reaches a mean test AUC of $0.8002 \pm 0.0052$, compared with $0.7922 \pm 0.0138$ for MERA($U=I$), and the low-false-positive-rate operating point changes from $1.10\times10^{-3}$ to $1.37\times10^{-3}$ in mean $\mathrm{TPR}@10^{-3}$. Using matched evaluation seeds, the mean differences are $\Delta\mathrm{AUC}=0.0081$ and $\Delta\mathrm{TPR}@10^{-3}=2.71\times10^{-4}$. The uncertainties on the AUC values overlap, however, so this finite-seed comparison should be interpreted as a positive mean shift rather than a statistically resolved performance improvement.

To determine whether the learned disentanglers nevertheless perform a nontrivial transformation, we also measured their distance from the identity at the end of training. Averaged first over the 45 encoder tensors in each model and then over the same eight evaluation seeds, we find
\begin{equation}
\left\langle \|U-I_6\|_F \right\rangle = 3.898 \pm 0.026,
\qquad
\left\langle \frac{\|U-I_6\|_F}{\sqrt{6}} \right\rangle = 1.591 \pm 0.011,
\end{equation}
where the quoted spread is the standard deviation across seeds. Thus, the trained local mixing maps do not remain close to the identity: they learn substantial transformations even though the present AUC comparison alone does not establish a statistically significant performance gain. As a diagnostic of the soft structural constraints, the corresponding residuals are
\begin{equation}
\left\langle \|U^\top U-I_6\|_F \right\rangle = 1.623 \pm 0.110,
\qquad
\left\langle \|W W^\top-I_3\|_F \right\rangle = 0.228 \pm 0.034.
\end{equation}
The normalized residuals, obtained by dividing by $\sqrt{6}$ and $\sqrt{3}$, are $0.662\pm0.045$ and $0.131\pm0.020$, respectively. These nonzero values quantify the fact that the penalties are soft: the trained model retains the intended MERA-inspired structure but is not an exactly orthogonal/isometric canonical MERA.

This is the regime in which the MERA-inspired architectural motivation is strongest, and it is precisely there that the local-mixing ablation tests it. The result does not mean that every additional tensor must help in every possible setting. It says something more specific: when the implemented anomaly detector operates in the strongest-compression regime and the locality-preserving input representation is used, the trainable local mixing maps move substantially away from the identity and are accompanied by a positive mean performance shift. The present seed sample does not resolve that shift statistically, but the structural diagnostic shows directly that this component of the implemented MERA-inspired neural architecture is not merely an identity-like or inactive addition.

\begin{table}[t]
\centering
\small
\caption{Focused multi-seed disentangler ablation at the strongest-compression setting $B=8$. MERA($U=I$) denotes the identity-disentangler limit of the same hierarchy, in which the local disentangling maps are fixed rather than learned. Both models are evaluated on the same eight seeds after separate retuning at $B=8$. The last row gives the mean matched-seed gain of the full MERA over MERA($U=I$).}
\label{tab:mera_disentangler_ablation}
\begin{tabular*}{\textwidth}{@{\extracolsep{\fill}}lccc@{}}
\toprule
Model & Params & Test AUC & $\mathrm{TPR}@10^{-3}$ \\
\midrule
MERA         & 5021 & $0.8002 \pm 0.0052$ & $1.37\times10^{-3}$ \\
MERA($U=I$) & 1781 & $0.7922 \pm 0.0138$ & $1.10\times10^{-3}$ \\
\midrule
Gain of full MERA & --- & $+0.0081$ & $+2.71\times10^{-4}$ \\
\bottomrule
\end{tabular*}
\end{table}

\section{Summary and Conclusions}
\label{sec:conclusions}

We set out to test two related questions. First, can a MERA-inspired autoencoder serve as an effective background-only anomaly detector for jets when applied directly to an ordered constituent representation? Second, are the architectural ingredients that motivate MERA in the first place, namely locality-aligned hierarchical coarse-graining and local correlation-reorganisation operations, supported by the data? To answer those questions, we construct the MERA-inspired neural encoder--decoder defined in \Sec{sec:mera_arch} and benchmark it on the Top Quark Tagging Reference Dataset against a dense autoencoder, the corresponding identity-mixing TTN limit, and classical unsupervised baselines. The numerical conclusions of this study apply to that implemented architecture, including its within-pair connectivity, nonlinear activations, untied decoder, and soft structural constraints; they are not claims about canonical MERA in general.

The nominal benchmark gives a clear positive result for the implemented MERA-inspired construction. We find that, for the locality-preserving representation used throughout the paper, this model achieves the largest mean test AUC at all three latent dimensions considered, with values of $0.7959$, $0.7885$, and $0.7963$ at $B=8,16,32$, respectively. The corresponding dense-autoencoder values are $0.7616$, $0.7487$, and $0.7550$, so the mean AUC gains are about $0.034$, $0.040$, and $0.041$. At the same time, the MERA-inspired models use only about one third of the trainable parameters of the dense autoencoder, with roughly $5.0$--$5.5\times10^3$ parameters versus $1.68$--$1.80\times10^4$ for AE. The nominal benchmark therefore shows that this particular multiscale neural architecture is both stronger in mean AUC and more economical than the standard nonlinear reconstruction baseline in the setting studied here.

The targeted follow-up studies then clarify why this happens. We showed that the locality-preserving ordering reduces the mean adjacent constituent separation to $\langle\Delta R\rangle=0.131$, compared with $0.361$ for $p_T$ ordering and $0.397\pm0.004$ for random permutations, and that it also gives the largest retained-variance fractions in the first two local compression layers, namely $0.869$ and $0.920$. This is precisely the regime in which the local MERA tensors act. Consistently, the completed ordering study shows that the same locality-preserving representation gives the strongest MERA performance at every latent dimension, with mean AUC values of $0.780$, $0.777$, and $0.774$, compared with $0.745$, $0.745$, and $0.756$ for $p_T$ ordering and $0.702$, $0.698$, and $0.702$ for random orderings. Through this analysis, we showed that the ordering is not a minor preprocessing detail: it is part of the mechanism that allows the hierarchical compression to align with the physical organisation of the jet.

The disentangler study provides a more specific but equally important result. In the strongest-compression regime, $B=8$, the full MERA exhibits a positive mean shift relative to the identity-disentangler limit MERA($U=I$), which operationally mimics the corresponding TTN by setting each disentangler to the identity while leaving the rest of the algorithm unchanged. The full MERA reaches $0.8002 \pm 0.0052$ in mean test AUC, compared with $0.7922 \pm 0.0138$ for MERA($U=I$), while the mean $\mathrm{TPR}@10^{-3}$ changes favourably from $1.10\times10^{-3}$ to $1.37\times10^{-3}$. This is the regime in which the mathematical argument of \Sec{sec:mera_construction} is most relevant: if the compression bottleneck is tight, then the ability of the disentanglers to rotate short-range correlations into the subspace retained by the isometries should matter most. The trained maps move substantially away from the identity and are accompanied by favourable mean shifts in both reported performance measures. The ablation therefore establishes that the disentangling step is an active, nontrivial component of the architecture, although the present eight-seed sample does not statistically resolve its performance advantage.

Thus, we established that the implemented MERA-inspired neural autoencoder is a viable and parameter-efficient architecture for collider anomaly detection on ordered jet constituents; that locality-aware preprocessing is a substantive part of why this multiscale model works; and that, in the regime where the hierarchy is forced to compress most aggressively, its trainable local mixing maps are demonstrably active and accompany a positive mean performance shift relative to the corresponding identity-map limit. Taken together, these results provide a concrete empirical case that MERA-inspired, multiscale, locality-aware compression can be a useful inductive bias for jet anomaly detection. 

\acknowledgments
E.G. was supported by the Scientific and Technological Research Council of Türkiye (TÜBİTAK) under the 2219 International Postdoctoral Research Fellowship Program (Grant No.1059B192400291).

\bibliographystyle{inspire}
\bibliography{apssamp}

@PREAMBLE{
 "\providecommand{\noopsort}[1]{}" 
 # "\providecommand{\singleletter}[1]{#1}%" 
}

@article{Fraser:2021lxm,
    author = "Fraser, Katherine and Homiller, Samuel and Mishra, Rashmish K. and Ostdiek, Bryan and Schwartz, Matthew D.",
    title = "{Challenges for unsupervised anomaly detection in particle physics}",
    eprint = "2110.06948",
    archivePrefix = "arXiv",
    primaryClass = "hep-ph",
    doi = "10.1007/JHEP03(2022)066",
    journal = "JHEP",
    volume = "03",
    pages = "066",
    year = "2022"
}

@article{Belis:2024anomaly,
    author = "Belis, Vasilis and Odagiu, Patrick and Aarrestad, Thea Klaeboe",
    title = "{Machine learning for anomaly detection in particle physics}",
    eprint = "2312.14190",
    archivePrefix = "arXiv",
    primaryClass = "physics.data-an",
    doi = "10.1016/j.revip.2024.100091",
    journal = "Rev. Phys.",
    volume = "12",
    pages = "100091",
    year = "2024"
}

@article{Kasieczka:2021lhc,
    author = "Kasieczka, Gregor and others",
    title = "{The LHC Olympics 2020 a community challenge for anomaly detection in high energy physics}",
    eprint = "2101.08320",
    archivePrefix = "arXiv",
    primaryClass = "hep-ph",
    doi = "10.1088/1361-6633/ac36b9",
    journal = "Rept. Prog. Phys.",
    volume = "84",
    number = "12",
    pages = "124201",
    year = "2021"
}

@article{Heimel:2019qcd,
    author = "Heimel, Theo and Kasieczka, Gregor and Plehn, Tilman and Thompson, Jennifer M.",
    title = "{QCD or What?}",
    eprint = "1808.08979",
    archivePrefix = "arXiv",
    primaryClass = "hep-ph",
    doi = "10.21468/SciPostPhys.6.3.030",
    journal = "SciPost Phys.",
    volume = "6",
    number = "3",
    pages = "030",
    year = "2019"
}

@article{Farina:2020deepae,
    author = "Farina, Marco and Nakai, Yuichiro and Shih, David",
    title = "{Searching for New Physics with Deep Autoencoders}",
    eprint = "1808.08992",
    archivePrefix = "arXiv",
    primaryClass = "hep-ph",
    doi = "10.1103/PhysRevD.101.075021",
    journal = "Phys. Rev. D",
    volume = "101",
    number = "7",
    pages = "075021",
    year = "2020"
}

@article{Finke:2021autoencoders,
    author = "Finke, Thorben and Kr{\"a}mer, Michael and Morandini, Alessandro and M{\"u}ck, Alexander and Oleksiyuk, Ivan",
    title = "{Autoencoders for unsupervised anomaly detection in high energy physics}",
    eprint = "2104.09051",
    archivePrefix = "arXiv",
    primaryClass = "hep-ph",
    doi = "10.1007/JHEP06(2021)161",
    journal = "JHEP",
    volume = "06",
    pages = "161",
    year = "2021"
}

@article{Larkoski:2020jetsub,
    author = "Larkoski, Andrew J. and Moult, Ian and Nachman, Benjamin",
    title = "{Jet substructure at the Large Hadron Collider: a review of recent advances in theory and machine learning}",
    eprint = "1709.04464",
    archivePrefix = "arXiv",
    primaryClass = "hep-ph",
    doi = "10.1016/j.physrep.2019.11.001",
    journal = "Phys. Rept.",
    volume = "841",
    pages = "1--63",
    year = "2020"
}

@article{Kasieczka:2019landscape,
    author = "Kasieczka, Gregor and others",
    title = "{The Machine Learning landscape of top taggers}",
    eprint = "1902.09914",
    archivePrefix = "arXiv",
    primaryClass = "hep-ph",
    doi = "10.21468/SciPostPhys.7.1.014",
    journal = "SciPost Phys.",
    volume = "7",
    number = "1",
    pages = "014",
    year = "2019"
}

@article{Vidal:2007er,
    author = "Vidal, Guifr{\'e}",
    title = "{Entanglement renormalization}",
    doi = "10.1103/PhysRevLett.99.220405",
    journal = "Phys. Rev. Lett.",
    volume = "99",
    number = "22",
    pages = "220405",
    year = "2007"
}

@article{Vidal:2008mera,
    author = "Vidal, Guifr{\'e}",
    title = "{Class of quantum many-body states that can be efficiently simulated}",
    doi = "10.1103/PhysRevLett.101.110501",
    journal = "Phys. Rev. Lett.",
    volume = "101",
    number = "11",
    pages = "110501",
    year = "2008"
}

@article{Stoudenmire:2016tnml,
    author = "Stoudenmire, E. Miles and Schwab, David J.",
    title = "{Supervised Learning with Tensor Networks}",
    eprint = "1605.05775",
    archivePrefix = "arXiv",
    primaryClass = "cs.LG",
    journal = "Adv. Neural Inf. Process. Syst.",
    volume = "29",
    pages = "4799--4807",
    year = "2016"
}

@article{Orus:2014intro,
    author = "Or{\'u}s, Rom{\'a}n",
    title = "{A practical introduction to tensor networks: Matrix product states and projected entangled pair states}",
    eprint = "1306.2164",
    archivePrefix = "arXiv",
    primaryClass = "cond-mat.str-el",
    doi = "10.1016/j.aop.2014.06.013",
    journal = "Annals Phys.",
    volume = "349",
    pages = "117--158",
    year = "2014"
}

@article{Cichocki:2017tnreview,
    author = "Cichocki, Andrzej and Phan, Anh-Huy and Zhao, Qibin and Lee, Namgil and Oseledets, Ivan and Sugiyama, Masashi and Mandic, Danilo P.",
    title = "{Tensor Networks for Dimensionality Reduction and Large-scale Optimization: Part 2 Applications and Future Perspectives}",
    doi = "10.1561/2200000067",
    journal = "Found. Trends Mach. Learn.",
    volume = "9",
    number = "4-5",
    pages = "431--673",
    year = "2017"
}

@article{Stoudenmire:2018multiscale,
    author = "Stoudenmire, E. Miles",
    title = "{Learning relevant features of data with multi-scale tensor networks}",
    doi = "10.1088/2058-9565/aaba1a",
    journal = "Quantum Sci. Technol.",
    volume = "3",
    number = "3",
    pages = "034003",
    year = "2018"
}

@article{Reyes:2021multiscale,
    author = "Reyes, J. A. and Stoudenmire, E. M.",
    title = "{Multi-scale tensor network architecture for machine learning}",
    doi = "10.1088/2632-2153/abffe8",
    journal = "Mach. Learn. Sci. Technol.",
    volume = "2",
    number = "3",
    pages = "035036",
    year = "2021"
}

@article{Liu:2019unitarytn,
    author = "Liu, Ding and Ran, Shi-Ju and Wittek, Peter and Peng, Cheng and Bl{\'a}zquez Garc{\'i}a, Raul and Su, Gang and Lewenstein, Maciej",
    title = "{Machine learning by unitary tensor network of hierarchical tree structure}",
    eprint = "1710.04833",
    archivePrefix = "arXiv",
    primaryClass = "stat.ML",
    doi = "10.1088/1367-2630/ab31ef",
    journal = "New J. Phys.",
    volume = "21",
    number = "7",
    pages = "073059",
    year = "2019"
}

@article{Han:2018mpsgen,
    author = "Han, Zhao-Yu and Wang, Jun and Fan, Heng and Wang, Lei and Zhang, Pan",
    title = "{Unsupervised Generative Modeling Using Matrix Product States}",
    doi = "10.1103/PhysRevX.8.031012",
    journal = "Phys. Rev. X",
    volume = "8",
    number = "3",
    pages = "031012",
    year = "2018"
}

@article{Cheng:2019ttngen,
    author = "Cheng, Song and Wang, Lei and Xiang, Tao and Zhang, Pan",
    title = "{Tree tensor networks for generative modeling}",
    eprint = "1901.02217",
    archivePrefix = "arXiv",
    primaryClass = "cond-mat.str-el",
    doi = "10.1103/PhysRevB.99.155131",
    journal = "Phys. Rev. B",
    volume = "99",
    number = "15",
    pages = "155131",
    year = "2019"
}

@article{Batselier:2021meracle,
    author = "Batselier, Kim and Cichocki, Andrzej and Wong, Ngai",
    title = "{MERACLE: Constructive Layer-Wise Conversion of a Tensor Train into a MERA}",
    doi = "10.1007/s42967-020-00090-6",
    journal = "Commun. Appl. Math. Comput.",
    volume = "3",
    number = "2",
    pages = "257--279",
    year = "2021"
}

@article{Sjostrand:2015pythia,
    author = "Sj{\"o}strand, Torbj{\"o}rn and Ask, Stefan and Christiansen, Jesper R. and Corke, Richard and Desai, Nishita and Ilten, Philip and Mrenna, Stephen and Prestel, Stefan and Rasmussen, Christine O. and Skands, Peter Z.",
    title = "{An Introduction to PYTHIA 8.2}",
    eprint = "1410.3012",
    archivePrefix = "arXiv",
    primaryClass = "hep-ph",
    doi = "10.1016/j.cpc.2015.01.024",
    journal = "Comput. Phys. Commun.",
    volume = "191",
    pages = "159--177",
    year = "2015"
}

@article{deFavereau:2014delphes,
    author = "de Favereau, J. and others",
    collaboration = "DELPHES 3",
    title = "{DELPHES 3, A modular framework for fast simulation of a generic collider experiment}",
    eprint = "1307.6346",
    archivePrefix = "arXiv",
    primaryClass = "hep-ex",
    doi = "10.1007/JHEP02(2014)057",
    journal = "JHEP",
    volume = "02",
    pages = "057",
    year = "2014"
}

@article{Cacciari:2008antikt,
    author = "Cacciari, Matteo and Salam, Gavin P. and Soyez, Gregory",
    title = "{The anti-$k_t$ jet clustering algorithm}",
    eprint = "0802.1189",
    archivePrefix = "arXiv",
    primaryClass = "hep-ph",
    doi = "10.1088/1126-6708/2008/04/063",
    journal = "JHEP",
    volume = "04",
    pages = "063",
    year = "2008"
}

@article{Hinton:2006autoencoder,
    author = "Hinton, Geoffrey E. and Salakhutdinov, Ruslan R.",
    title = "{Reducing the Dimensionality of Data with Neural Networks}",
    doi = "10.1126/science.1127647",
    journal = "Science",
    volume = "313",
    number = "5786",
    pages = "504--507",
    year = "2006"
}

@article{Pearson:1901pca,
    author = "Pearson, Karl",
    title = "{LIII. On lines and planes of closest fit to systems of points in space}",
    doi = "10.1080/14786440109462720",
    journal = "Philos. Mag.",
    volume = "2",
    number = "11",
    pages = "559--572",
    year = "1901"
}

@article{Hotelling:1933pca,
    author = "Hotelling, Harold",
    title = "{Analysis of a complex of statistical variables into principal components}",
    doi = "10.1037/h0071325",
    journal = "J. Educ. Psychol.",
    volume = "24",
    number = "6",
    pages = "417--441",
    year = "1933"
}

@article{Mahalanobis:1936distance,
    author = "Mahalanobis, P. C.",
    title = "{On the generalised distance in statistics}",
    journal = "Proc. Natl. Inst. Sci. India",
    volume = "2",
    pages = "49--55",
    year = "1936"
}

@inproceedings{Liu:2008iforest,
    author = "Liu, Fei Tony and Ting, Kai Ming and Zhou, Zhi-Hua",
    title = "{Isolation Forest}",
    booktitle = "{2008 Eighth IEEE International Conference on Data Mining}",
    doi = "10.1109/ICDM.2008.17",
    pages = "413--422",
    year = "2008"
}

@article{Ngairangbam:2022qaead,
    author = "Ngairangbam, Vishal S. and Spannowsky, Michael and Takeuchi, Michihisa",
    title = "{Anomaly detection in high-energy physics using a quantum autoencoder}",
    doi = "10.1103/PhysRevD.105.095004",
    journal = "Phys. Rev. D",
    volume = "105",
    number = "9",
    pages = "095004",
    year = "2022"
}

@article{Puljak:2025tnad,
    author = "Puljak, Ema and Pierini, Maurizio and Garcia-Saez, Artur",
    title = "{Tensor Network for Anomaly Detection in the Latent Space of Proton Collision Events at the LHC}",
    eprint = "2506.00102",
    archivePrefix = "arXiv",
    primaryClass = "stat.ML",
    doi = "10.1088/2632-2153/ae0243",
    journal = "Mach. Learn. Sci. Technol.",
    volume = "6",
    number = "4",
    pages = "045001",
    year = "2025"
}

@article{Araz:2022zxk,
    author = "Araz, Jack Y. and Spannowsky, Michael",
    title = "{Quantum-probabilistic Hamiltonian learning for generative modeling and anomaly detection}",
    eprint = "2211.03803",
    archivePrefix = "arXiv",
    primaryClass = "quant-ph",
    reportNumber = "IPPP/22/77",
    doi = "10.1103/PhysRevA.108.062422",
    journal = "Phys. Rev. A",
    volume = "108",
    number = "6",
    pages = "062422",
    year = "2023"
}

@article{Araz:2022haf,
    author = "Araz, Jack Y. and Spannowsky, Michael",
    title = "{Classical versus quantum: Comparing tensor-network-based quantum circuits on Large Hadron Collider data}",
    eprint = "2202.10471",
    archivePrefix = "arXiv",
    primaryClass = "quant-ph",
    reportNumber = "IPPP/22/06",
    doi = "10.1103/PhysRevA.106.062423",
    journal = "Phys. Rev. A",
    volume = "106",
    number = "6",
    pages = "062423",
    year = "2022"
}

@article{Araz:2021zwu,
    author = "Araz, Jack Y. and Spannowsky, Michael",
    title = "{Quantum-inspired event reconstruction with Tensor Networks: Matrix Product States}",
    eprint = "2106.08334",
    archivePrefix = "arXiv",
    primaryClass = "hep-ph",
    reportNumber = "IPPP/20/114",
    doi = "10.1007/JHEP08(2021)112",
    journal = "JHEP",
    volume = "08",
    pages = "112",
    year = "2021"
}

@article{Collins:2021weakunsup,
    author = "Collins, Jack H. and Martin-Ramiro, Pablo and Nachman, Benjamin and Shih, David",
    title = "{Comparing weak- and unsupervised methods for resonant anomaly detection}",
    journal = "Eur. Phys. J. C",
    volume = "81",
    pages = "617",
    year = "2021",
    doi = "10.1140/epjc/s10052-021-09389-x"
}

@article{Barron:2021suep,
    author = "Barron, Jared and Curtin, David and Kasieczka, Gregor and Plehn, Tilman and Spourdalakis, Aris",
    title = "{Unsupervised hadronic SUEP at the LHC}",
    journal = "JHEP",
    volume = "12",
    pages = "129",
    year = "2021",
    doi = "10.1007/JHEP12(2021)129"
}

@article{Blance:2019adversarialae,
    author = "Blance, Andrew and Spannowsky, Michael and Waite, Philip",
    title = "{Adversarially-trained autoencoders for robust unsupervised new physics searches}",
    journal = "JHEP",
    volume = "10",
    pages = "047",
    year = "2019",
    doi = "10.1007/JHEP10(2019)047"
}

@article{Atkinson:2022ircgraphae,
    author = "Atkinson, Oliver and Bhardwaj, Akanksha and Englert, Christoph and Konar, Partha and Ngairangbam, Vishal S. and Spannowsky, Michael",
    title = "{IRC-Safe Graph Autoencoder for Unsupervised Anomaly Detection}",
    journal = "Front. Artif. Intell.",
    volume = "5",
    pages = "943135",
    year = "2022",
    doi = "10.3389/frai.2022.943135"
}

@article{Dillon:2023normalizedae,
    author = "Dillon, Barry M. and Favaro, Luigi and Plehn, Tilman and Sorrenson, Peter and Kr{\"a}mer, Michael",
    title = "{A normalized autoencoder for LHC triggers}",
    journal = "SciPost Phys. Core",
    volume = "6",
    pages = "074",
    year = "2023",
    doi = "10.21468/SciPostPhysCore.6.4.074"
}

@article{Finke:2024treebased,
    author = "Finke, Thorben and Hein, Marie and Kasieczka, Gregor and Kr{\"a}mer, Michael and M{\"u}ck, Alexander and Prangchaikul, Parada and Quadfasel, Tobias and Shih, David and Sommerhalder, Manuel",
    title = "{Tree-based algorithms for weakly supervised anomaly detection}",
    journal = "Phys. Rev. D",
    volume = "109",
    pages = "034033",
    year = "2024",
    doi = "10.1103/PhysRevD.109.034033"
}

@misc{Kasieczka:2019topdataset,
    author = "Kasieczka, Gregor and Plehn, Tilman and Thompson, Jennifer and Russel, Michael",
    title = "{Top Quark Tagging Reference Dataset}",
    publisher = "Zenodo",
    year = "2019",
    month = "3",
    doi = "10.5281/zenodo.2603256",
    url = "https://doi.org/10.5281/zenodo.2603256",
    note = "Version v0 (2018\_03\_27)"
}

@article{Edelman:1998orthogonality,
    author = "Edelman, Alan and Arias, Tom\'as A. and Smith, Steven T.",
    title = "{The Geometry of Algorithms with Orthogonality Constraints}",
    journal = "SIAM J. Matrix Anal. Appl.",
    volume = "20",
    number = "2",
    pages = "303--353",
    year = "1998",
    doi = "10.1137/S0895479895290954"
}

@book{Absil:2008manifolds,
    author = "Absil, Pierre-Antoine and Mahony, Robert and Sepulchre, Rodolphe",
    title = "{Optimization Algorithms on Matrix Manifolds}",
    publisher = "Princeton University Press",
    address = "Princeton",
    year = "2008",
    doi = "10.1515/9781400830244"
}

@article{DeLathauwer:2000hosvd,
    author = "De Lathauwer, Lieven and De Moor, Bart and Vandewalle, Joos",
    title = "{A Multilinear Singular Value Decomposition}",
    journal = "SIAM J. Matrix Anal. Appl.",
    volume = "21",
    number = "4",
    pages = "1253--1278",
    year = "2000",
    doi = "10.1137/S0895479896305696"
}

@article{Schonemann:1966procrustes,
    author = "Sch{\"o}nemann, Peter H.",
    title = "{A Generalized Solution of the Orthogonal Procrustes Problem}",
    journal = "Psychometrika",
    volume = "31",
    number = "1",
    pages = "1--10",
    year = "1966",
    doi = "10.1007/BF02289451"
}

@inproceedings{Kingma:2014vow,
    author = "Kingma, Diederik P. and Ba, Jimmy",
    title = "{Adam: A Method for Stochastic Optimization}",
    booktitle = "{International Conference on Learning Representations}",
    eprint = "1412.6980",
    archivePrefix = "arXiv",
    primaryClass = "cs.LG",
    month = "12",
    year = "2014"
}
\end{document}